\documentclass[12pt]{article}
\usepackage{xcolor}
\usepackage{graphicx}
\usepackage{multirow}
\usepackage{array}
\usepackage{cite}
\usepackage{amssymb}
\usepackage{amsmath}
\usepackage{pifont}
\usepackage[colorlinks,linkcolor=red,anchorcolor=black,citecolor=green]{hyperref}
\usepackage{mathrsfs}
\usepackage{soul}
\sethlcolor{lightgray}
\graphicspath{{./figs/}}
\allowdisplaybreaks \allowdisplaybreaks[2]

\begin{document}

\title{\hfill ~\\[-30mm] \hfill\mbox{\small USTC-ICTS-18-05}\\[10mm]
        \textbf{\large Systematic Analysis of Dirac Neutrino Masses at Dimension Five}}

\date{}

\author{\\[1mm]Chang-Yuan Yao\footnote{Email: {\tt
phyman@mail.ustc.edu.cn}}~,~Gui-Jun Ding\footnote{Email: {\tt
dinggj@ustc.edu.cn}}\\ \\
\it{\small Interdisciplinary Center for Theoretical Study and  Department of
Modern Physics, }\\
\it{\small University of Science and
    Technology of China, Hefei, Anhui 230026, China}\\[4mm] }
\maketitle

\begin{abstract}
We perform a systematic study of the Dirac neutrino masses which arise from a dimension five effective operator with a singlet scalar. We identify all possible realizations of this operator at tree level and one-loop level. The corresponding predictions for the particle content and neutrino mass matrix are presented. We add a $Z_{2}$ symmetry to forbid the renormalizable Yukawa coupling, and non-abelian discrete flavor symmetry can be used to forbid tree level diagrams in non-genuine one-loop models. The electrically neutral particle mediating the loop diagram could be dark matter candidate. We give the possible dark matter particles and the restrictions on the parameter $\alpha$ for each model.
\end{abstract}
\thispagestyle{empty}
\vfill

\newpage
\section{\label{sec:intro}Introduction}

Neutrino oscillation experiments have made great progress in past decades, the three neutrino mixing angles and the mass squared differences have been measured with high precision. It has been firmly established that neutrinos have mass~\cite{Kajita:2016cak,McDonald:2016ixn}. The existence of neutrino masses definitely requires new physics beyond standard model (SM). So far we still have no clue about the nature of massive neutrinos, which could be Dirac or Majorana. The only feasible experiments having the potential of establishing that neutrinos are Majorana particles are the experiments searching for neutrinoless double beta decay. However, this rare process has not yet been observed despite great efforts over several decades. It is usually assumed that neutrinos are Majorana particles,
the Majorana neutrino mass models have been extensively studied and systematically classified in the literature~\cite{Bonnet:2012kz,Sierra:2014rxa,Simoes:2017kqb,Cepedello:2017eqf,Farzan:2012ev}.
However, one should keep in mind that the theoretical
assumption of Majorana neutrinos and the resulting lepton number violation have not been confirmed by any experiments. Consequently the possibility of Dirac neutrinos cannot be discounted. If neutrinos are Dirac particles, in the same way as quarks and charged leptons the neutrino masses are generated by the following Yukawa interaction term
\begin{equation}
\label{eq:lag_d4}\mathscr{L}_4^D=-y_{\alpha\beta}\overline{\ell_{L\alpha}}\widetilde{H}\nu_{R\beta}+\text{H.c.}\,,
\end{equation}
where $\ell_L=(\nu_{L},l_{L})^T$ is the left-handed lepton doublet,
$H=(H^+,H^0)^T$ is the Higgs doublet with
$\widetilde{H}=i\sigma_2H^{*}$, and $\nu_{R}$ denotes the right-handed
neutrino fields. Since the neutrino mass is tiny of order eV which is much smaller than the vacuum expectation value of the Higgs field $\langle H^{0}\rangle\simeq174$ GeV, the Yukawa coupling should be quite small $y_{\alpha\beta}\sim\mathcal{O}(10^{-11})$. Such a small value is intrinsically unacceptable to many people. In addition, because $\nu_{R}$ is a standard model singlet, there is no symmetry which prevents it from having a large Majorana mass. As a result, the left-handed neutrinos would obtain an effective small Majorana mass from the seesaw mechanism~\cite{Minkowski:1977sc,Yanagida:1979as,GellMann:1980vs,Mohapatra:1979ia,Schechter:1980gr}.
Hence generally a global or  gauge symmetry is imposed in such a way that the lepton number is conserved in Dirac neutrino models, the Majorana mass term for $\nu_R$ will be forbidden, and the neutrinos are guaranteed to be Dirac particles.

It is well-known that tiny neutrino masses can be naturally explained by seesaw mechanism if neutrinos are Majorana particles. In a similar way, the seesaw mechanism can also allows for Dirac neutrinos. Three types of seesaw-like models have been constructed, and they are called
type-I~\cite{Roncadelli:1983ty,Ma:2015raa,Chulia:2016ngi,CentellesChulia:2017koy},
type-II~\cite{Gu:2006dc,Valle:2016kyz,Bonilla:2016zef,Bonilla:2017ekt}
and type-III~\cite{Gu:2016hxh} Dirac seesaw mechanisms. The following dimensional five effective operator is induced by the Dirac seesaw mechanism
\begin{equation}
\label{eq:lag_d5}\mathscr{L}_5^D=-\frac{g_{\alpha\beta}}{\Lambda}\overline{\ell_{L\alpha}}\widetilde{H}\nu_{R\beta}S+\text{H.c.}\,,
\end{equation}
where $S$ is a scalar singlet, and $\Lambda$ denotes the new physics
scale. After the spontaneous symmetry breaking, both $H$ and $S$ will
acquire vacuum expectation values, and the Dirac neutrino masses will
be generated by this dimensional five operator. Since $S$ is a standard model (SM) gauge singlet scalar, additional symmetry is required in order to guarantee that $\mathscr{L}_5^D$ gives the leading order contribution to the Dirac neutrino masses. We shall impose a simple $Z_2$ auxiliary symmetry in this work. We assume that both $S$ and $\nu_{R}$ are odd ($-$) under $Z_2$ while the SM particles are even ($+$) under $Z_2$. As a consequence, the renormalizable Yukawa coupling in Eq.~\eqref{eq:lag_d4} is automatically forbidden by this $Z_2$ symmetry. The neutrino Dirac masses can also be generated through radiative corrections~\cite{Mohapatra:1987nx,Gu:2007ug,Kanemura:2011jj,Farzan:2012sa,Ma:2016mwh,Wang:2016lve,Ma:2017kgb,Yao:2017vtm}.
Please see~\cite{Bonilla:2016diq,Borah:2017leo,Wang:2017mcy} for recent radiative Dirac neutrino mass models.

In the present work, we will study the dimension five operator in Eq.~\eqref{eq:lag_d5}, and we shall find out all possible ultraviolet completions of this operator at tree level and one-loop level. We shall identify all topologies and the corresponding models where the finite one-loop diagrams give the dominant contributions to neutrino masses and the tree level Dirac seesaw is absent. The additional fields and the prediction for neutrino mass matrix will be presented for each model.

The structure of this paper is as follows: we discuss the tree level and one-loop realizations of the dimension five Dirac neutrino mass operator in section~\ref{sec:treeloop}, we list all genuine diagrams and possible quantum number assignments for the new mediator fields. In section~\ref{sec:forbidtree}, we show that non-abelian discrete flavor symmetry can forbid tree level contributions to the neutrino masses in non-genuine one-loop models. We discuss the dark matter compatibility of the one-loop models for Dirac neutrino mass generation in section~\ref{sec:darkmatter}. We conclude in section~\ref{sec:conclusion}. We show all the integrals which are needed to calculate the neutrino masses in Appendix~\ref{sec:app_mass}. We provide detailed lists of the possible one-loop models for Dirac neutrino mass generation in Appendix~\ref{sec:app_table}, the possible field assignments and dark matter candidates are shown. We present the complete list of viable models that are contained within others in Appendix~\ref{sec:within}.

\section{\label{sec:treeloop}Systematic decomposition of tree and one-loop diagrams}

\begin{figure}[t!]
\centering
\begin{tabular}{cc}
\includegraphics[width=0.22\linewidth]{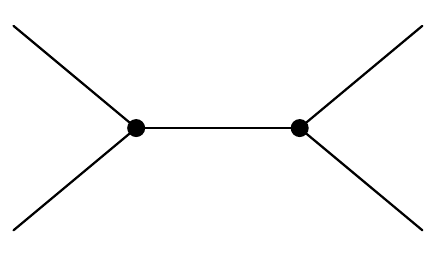} ~~~&~~~ \includegraphics[width=0.22\linewidth]{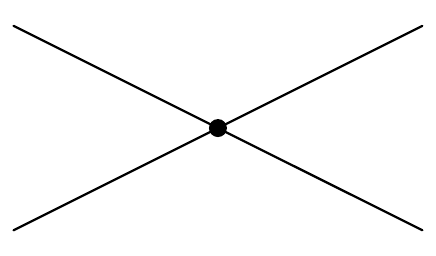}\\
 F1 & F2 \\
\end{tabular}
\caption{\label{fig:topfigtree}Topologies of tree level diagrams with four external legs.}
\end{figure}

\begin{figure}[t!]
\centering
\begin{tabular}{ccc}
\includegraphics[width=0.32\linewidth]{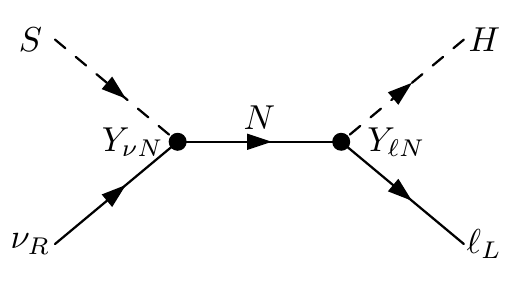} & \includegraphics[width=0.32\linewidth]{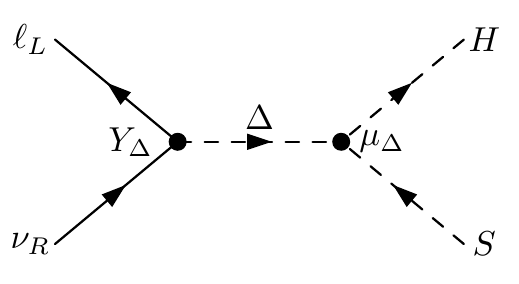} & \includegraphics[width=0.32\linewidth]{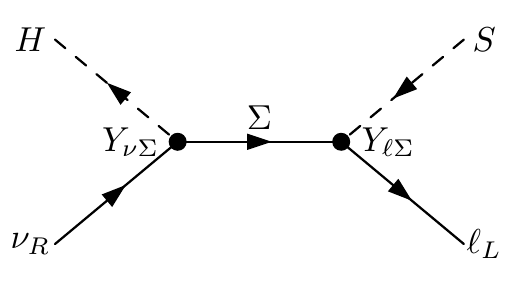}\\
type-I &  type-II  &  type-III
\end{tabular}
\caption{\label{fig:topfigtreeFS} The three realizations of the Dirac seesaw mechanism, known as type-I~(left), type-II~(middle) and type-III~(right) Dirac seesaw mechanism. The internal fields $N$, $\Delta$ and $\Sigma$ are singlet fermion, scalar doublet and fermionic doublet respectively. }
\end{figure}

In this section, we shall identify all possible tree level and one-loop ultraviolet completions of the dimensional five operator in Eq.~\eqref{eq:lag_d5}. At tree level one can construct two different topologies F1 and F2, as shown in figure~\ref{fig:topfigtree}. The topology F2 can not lead to any renormalizable model except that the four external legs are scalars. Consequently we will not analyze this topology further. For the topology F1, there are only three ways to de-construct the effective operator $\overline{\ell_{L\alpha}}\widetilde{H}\nu_{R\beta}S$ as shown in
figure~\ref{fig:topfigtreeFS}. They are known as type-I, type-II and
type-III Dirac seesaw respectively. The Dirac neutrino masses are generated via the introduction of a singlet fermion $N$, a scalar doublet $\Delta$ and a fermionic doublet $\Sigma$ respectively\footnote{Analogous to usual type-III seesaw models for the Majorana neutrinos, the type-III Dirac seesaw model in~\cite{Gu:2016hxh} extends the SM by the addition of two fermion triplets, a scalar triplet and a scalar doublet besides the singlet $S$. In contrast, the type-III Dirac seesaw is mediated by a doublet fermion in the present work. This simple Dirac seesaw model has not been previously proposed in the literature as far as we know.}. We list the quantum numbers of the mediators and neutrino masses for the three types of Dirac seesaw in table~\ref{tab:seesaw}. The quantum numbers of a field is given in a compact notation $X^{Z}_Y$, where $X$ refers to its $SU(2)_L$ transformation ($\mathbf{1}$ for singlet, $\mathbf{2}$ for doublet, and $\mathbf{3}$ for triplet), $Y$ denotes its hypercharge, and $Z$ stands for the $Z_2$ charges (``$+$'' for even and ``$-$'' for odd). In this work, we only consider the case that additional fields transforming as singlets, doublets, or triplets of $SU(2)_{L}$. The results for representation larger than $SU(2)_L$ triplets can be easily obtained in a similar way. The new fields are assumed to be either scalars or fermions, and the fermions should be vector-like to ensure anomaly cancellation. The diagrams with scalar or vector bosons are equivalent, and the resulting neutrino masses for the diagrams with vectors can be straightforwardly obtained from those of the diagrams with scalars. Furthermore, vector bosons are generally the gauge bosons of a certain gauge symmetry, their masses are generated via the spontaneous breaking of the gauge symmetry. As a result, the scalar sector of these models should be discussed carefully as well, the corresponding analysis is highly model dependent. Therefore we shall consider the scalar and fermion mediated models in this work.

\begin{table}[t!]
  \centering
\begin{tabular}{|c||c|c|}\hline\hline
   &   &  \\ [-0.16in]
  Dirac seesaw & mediator & $(m_{\nu})_{\alpha\beta}/(\langle H\rangle \langle S\rangle)$\\ 
    &   &  \\ [-0.16in]\hline
     &   &  \\ [-0.16in]
  type-I & $N\sim \mathbf{1}_0^{+}$ & $-\frac{(Y_{\ell N})_{\alpha i}(Y_{\nu N})_{i \beta}}{M_{N}^{(i)}}$ \\ 
     &   &  \\ [-0.16in] \hline
     &   &  \\ [-0.16in]
  type-II & $\Delta\sim \mathbf{2}_1^{-}$ & $-\frac{\mu_{\Delta} (Y_{\Delta})_{\alpha \beta}}{M_{\Delta}}$\\ 
     &   &  \\ [-0.16in] \hline
     &   &  \\ [-0.16in]
  type-III & $\Sigma\sim \mathbf{2}_{-1}^{-}$ & $-\frac{(Y_{\ell \Sigma})_{\alpha i}(Y_{\nu \Sigma})_{i \beta}}{M_{\Sigma}^{(i)}}$\\ [0.1in] \hline\hline
\end{tabular}
\caption{\label{tab:seesaw}The quantum number assignments of the
  mediators and the neutrino masses for three types of Dirac seesaw.}
\end{table}

\begin{figure}[!htb]
\centering
\includegraphics[width=0.65\linewidth]{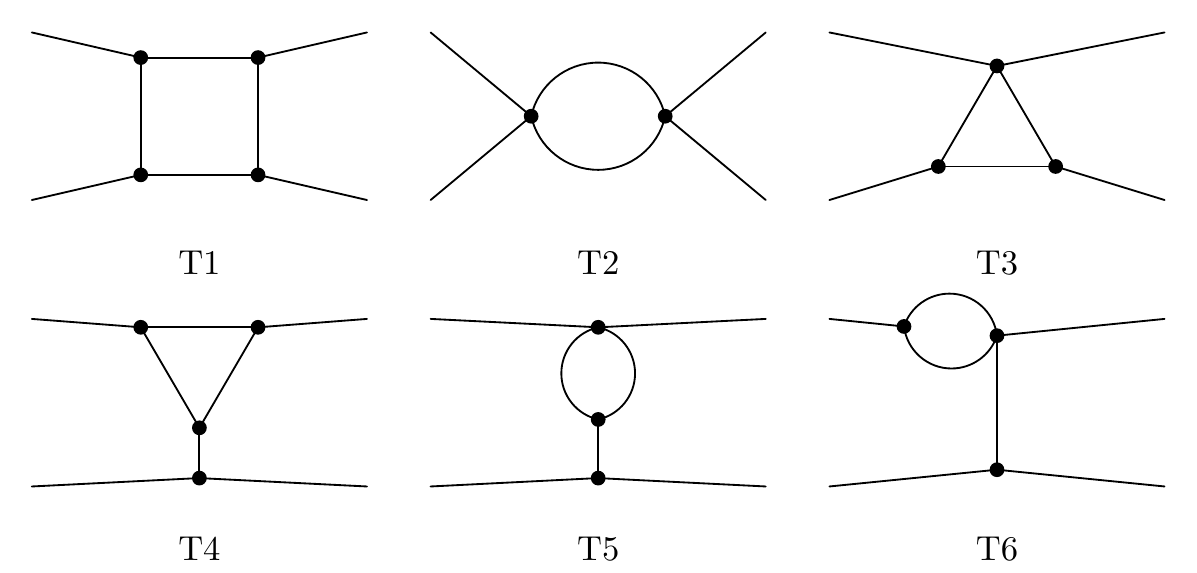}
\caption{\label{fig:topfigloop}Topologies of one-loop diagram with four external legs.}
\end{figure}

\begin{figure}[!htb]
\centering
\includegraphics[width=\linewidth]{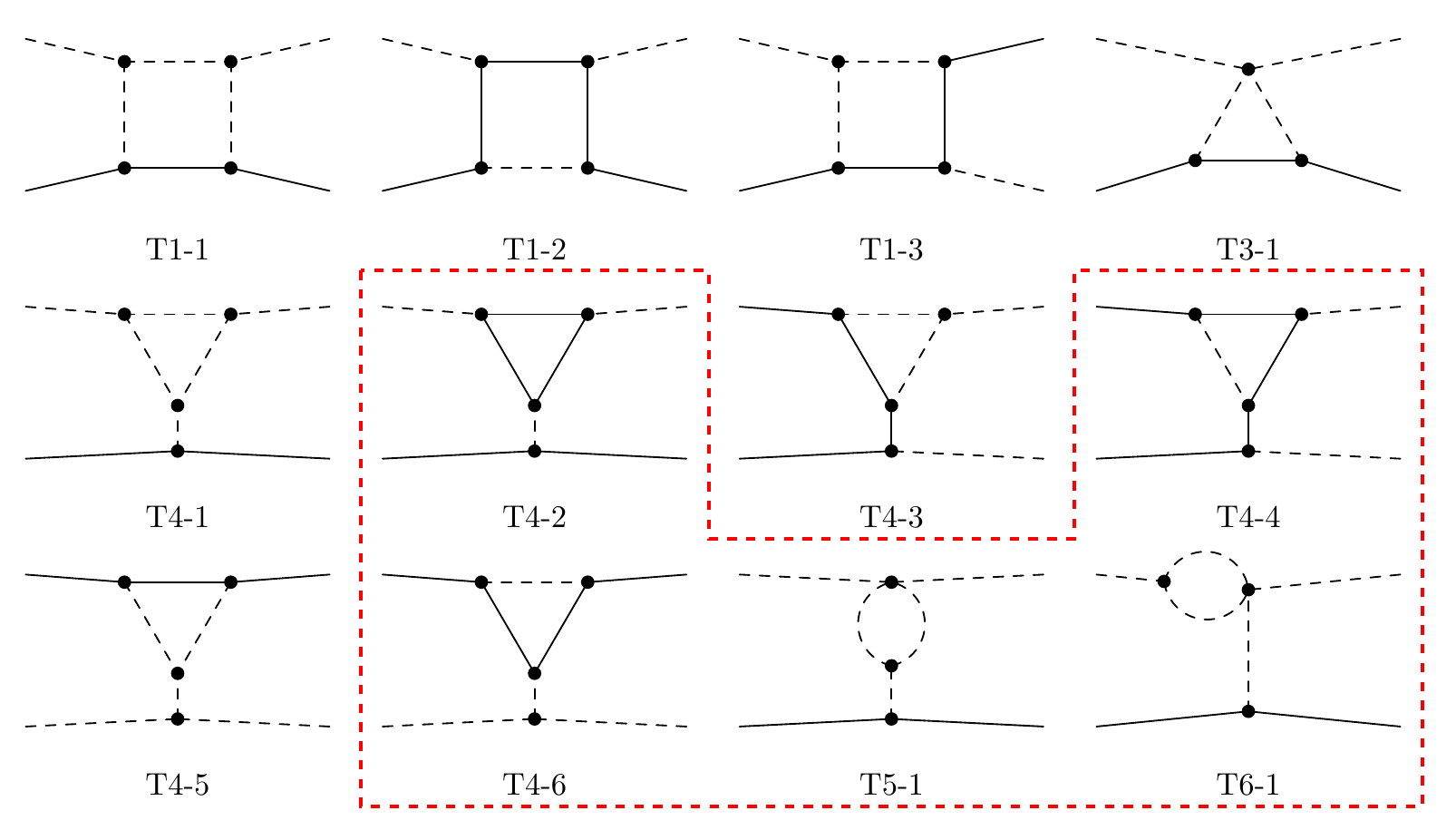}
\caption{\label{fig:topfigloopFS}The possible one-loop diagrams for topologies T1 to T6, where the dashed lines denote scalars, and the solid lines denote fermions. Here we require that two external legs are fermions and the remaining two external legs are scalars. The diagrams in the red dashed box are divergent.}
\end{figure}

\begin{figure}[!htb]
\centering
\includegraphics[width=0.8\linewidth]{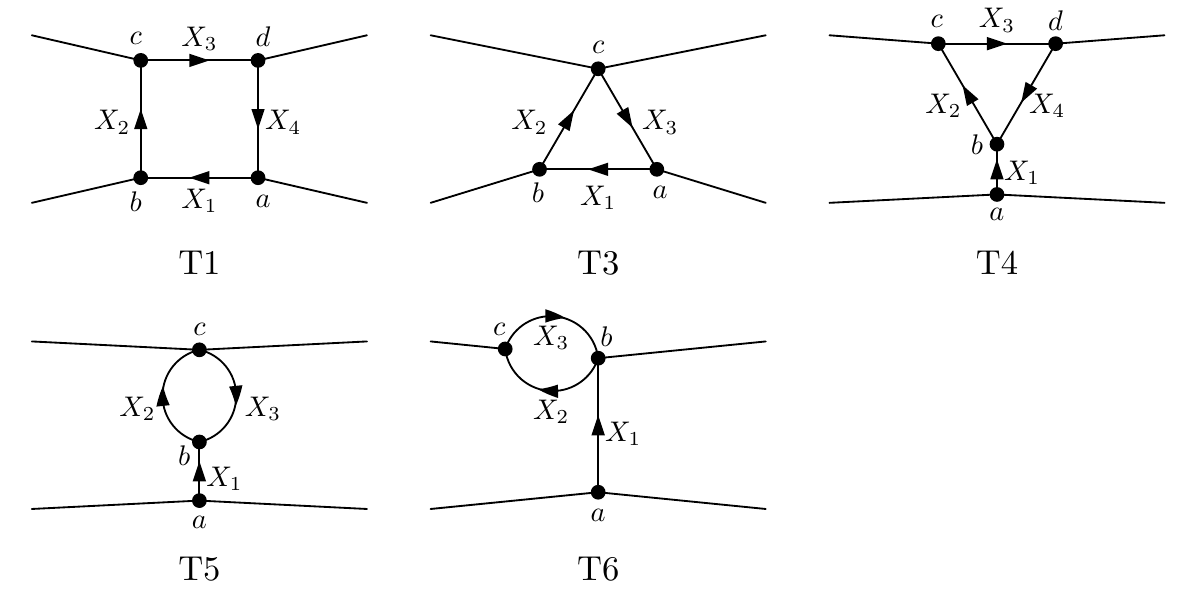}
\caption{\label{fig:topfigloopA}Symbolic internal field assignments
  and couplings for the different renormalizable one-loop diagrams.}
\end{figure}

\begin{figure}[!htb]
\centering
\includegraphics[width=0.8\linewidth]{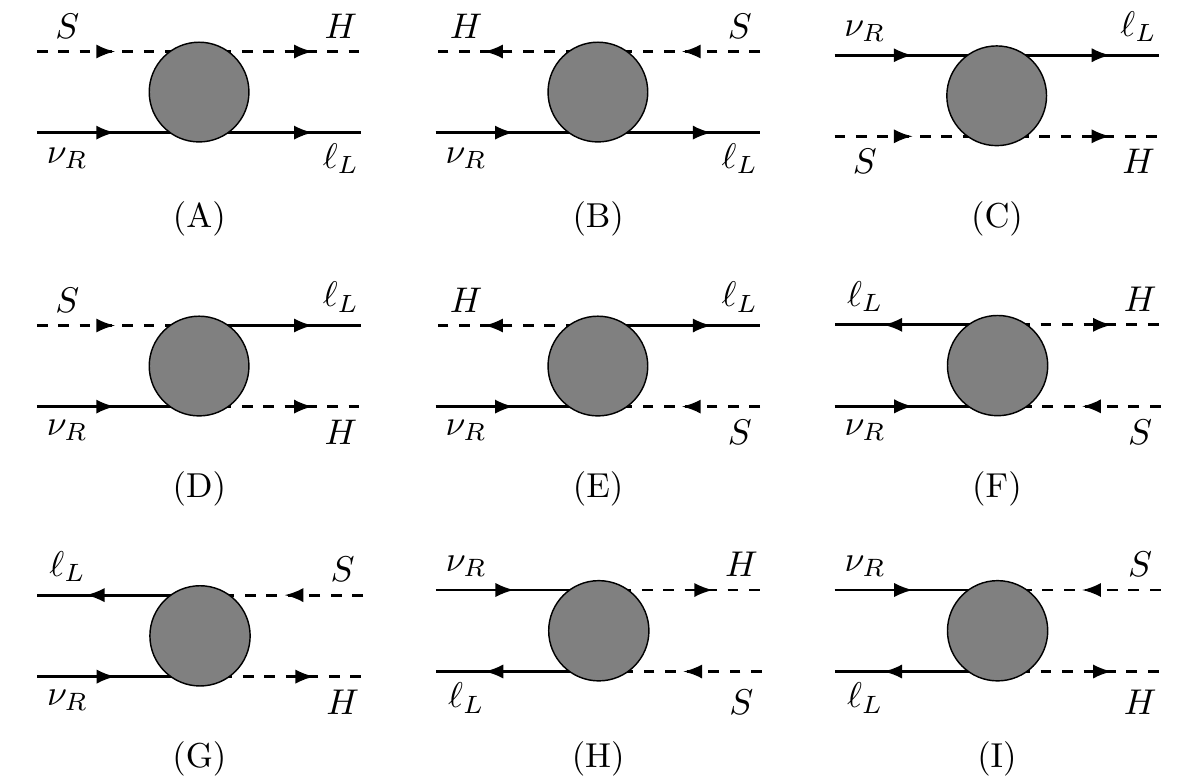}
\caption{\label{fig:eff_vertex}Possible external $\ell_{L},\,\nu_{R},\,H,\,S$ structures used to determine the standard model gauge charges of the mediators in the loop.}
\end{figure}

Following the diagram-based approach of
Refs.~\cite{Bonnet:2012kz,Sierra:2014rxa,Simoes:2017kqb,Yao:2017vtm},
we can find all possible one-loop realizations for the dimensional five Dirac neutrino mass operator in Eq.~\eqref{eq:lag_d5}. Firstly we use the powerful program \texttt{FeynArts}~\cite{Hahn:2000kx} to construct the one-loop topologies with four external legs, both self-energy and tadpole diagrams are excluded. In total there are only six possible topologies, as shown in figure~\ref{fig:topfigloop}. We shall discard the topology T2 because it leads to non-renormalizable operators. Then we specify the Lorentz nature (spinor or scalar) of each line. For each topology, we can use \texttt{FeynArts} to find out all possible scalar or fermionic assignments for lines. The possible one-loop diagrams arising from the renormalizable topologies are given in figure~\ref{fig:topfigloopFS}, where we denote the fermions with solid line, and scalars with dashed line. Since we are concerned with the one-loop realization of the effective operator in Eq.~\eqref{eq:lag_d5}, two external legs in figure~\ref{fig:topfigloopFS} are fermions and the remaining two external legs are scalars. The names of the internal fields and the couplings of the interaction vertices are defined in figure~\ref{fig:topfigloopA}. There are usually more than one possibilities of assigning the four external legs to the left-handed lepton doublet $\ell_L$, the right-handed neutrino singlet $\nu_R$, the Higgs doublets $H$ and the singlet scalar $S$. The patterns of the external legs are also related to the symmetry properties of the topologies given in figure~\ref{fig:topfigloopFS}. For example, the two scalar external legs of T3-1 are topologically equivalent, we only need to consider one possible assignment for them, since exchanging
these two legs doesn't give new diagram. We display the possible structures of the external fields in
figure~\ref{fig:eff_vertex}. Finally, we can construct all independent
diagrams by combining figure~\ref{fig:topfigloopFS}, figure~\ref{fig:topfigloopA} and figure~\ref{fig:eff_vertex}. The
generated one-loop Feynman diagrams are named as Ta-b-c, where ``a'' refers to the topology given in
figure~\ref{fig:topfigloopA}, ``b'' indicates the different choices of the
fermion and scalar lines in a given topology as shown in
figure~\ref{fig:topfigloopFS}, and ``c'' denotes the field assignments for the external legs given in figure~\ref{fig:eff_vertex}.

The loop integrals of the Feynman diagrams in figure~\ref{fig:topfigloopFS}
are either finite or divergent, and we collect all the divergent diagrams in a red dashed box. The divergence can be cancelled by the counter terms of the tree level Dirac seesaw. Hence the divergent diagrams can be regarded as corrections to the Dirac seesaws, and they are of no interest. For the remaining diagrams with finite loop integrals, we shall identify the diagrams for which both the renormalizable Yukawa coupling and the tree level Dirac seesaw are absent for certain quantum numbers of the mediators. The possible quantum numbers of the mediator fields and the predictions for neutrino masses are summarized in tables~\ref{tab:figT1}--\ref{tab:figT4}. In this work we consider the intermediate fields (scalars and fermions) are singlets, doublets or triplets of $SU(2)_{L}$. The results for larger representations can be easily obtained. We mention that all states are assumed to be color singlets for simplicity. Nevertheless, color charges can be trivially included since the lepton doublet $\ell_{L}$, right-handed neutrino $\nu_{R}$, Higgs doublet $H$ and the scalar singlet $S$ are all color singlets,

From tables~\ref{tab:figT1}--\ref{tab:figT4}, we see that the two topologies T1 and T3 can lead to genuine diagrams. Here we define genuine diagrams to be the diagrams for which the neutrino masses arise at one-loop level and the tree level contributions from Dirac seesaw are guaranteed to be absent. For each diagram, there are plenty of solutions for the quantum numbers of the extra fields beyond SM. For each given solution, the hypercharges of the fields are fixed up to a free parameter $\alpha$, and $\alpha$ should be an integer to avoid fractional charges. It is remarkable that three types of Dirac seesaw diagrams can be absent for certain values of $\alpha$, such that the neutrino masses are dominantly generated at one-loop level. In tables~\ref{tab:figT1}--\ref{tab:figT4}, we list the excluded values of $\alpha$ by the requirement that the tree level Dirac seesaw masses are absent. There are two possible ways to assign the $Z_2$ charges for the fields in the loop. The $Z_{2}$ charge is shown as "$\pm$" or "$\mp$" on the superscript of the quantum numbers. The "$+$" or "$-$" on the top is the first possible assignment for the $Z_{2}$ charges, we name it as $Z_{2}^{I}$, and another possible assignment at the bottom is called $Z_{2}^{II}$. We notice that sometimes different values of $\alpha$ are excluded for $Z_{2}^{I}$ and $Z_{2}^{II}$ to give neutrino masses at the one-loop level genuinely, as shown by two separate columns in these tables. For some models, the values of $\alpha$ excluded from the disappearance of tree level diagrams constitute a empty set $\varnothing$ or a universal set $\mathbb{U}$, it means that the tree level Dirac seesaw can't or can appear for any value of $\alpha$. If the excluded values of $\alpha$ constitute the
universal set $\mathbb{U}$ for both assignments $Z_{2}^{I}$ and $Z_{2}^{II}$, the tree level Dirac seesaw diagrams can not be avoided without additional symmetry. For completeness we present the assignments of the mediators for all possible diagrams in tables~\ref{tab:figT1}--\ref{tab:figT4}, including these non-genuine models.

\begin{figure}[t!]
\centering
\includegraphics[width=0.9\linewidth]{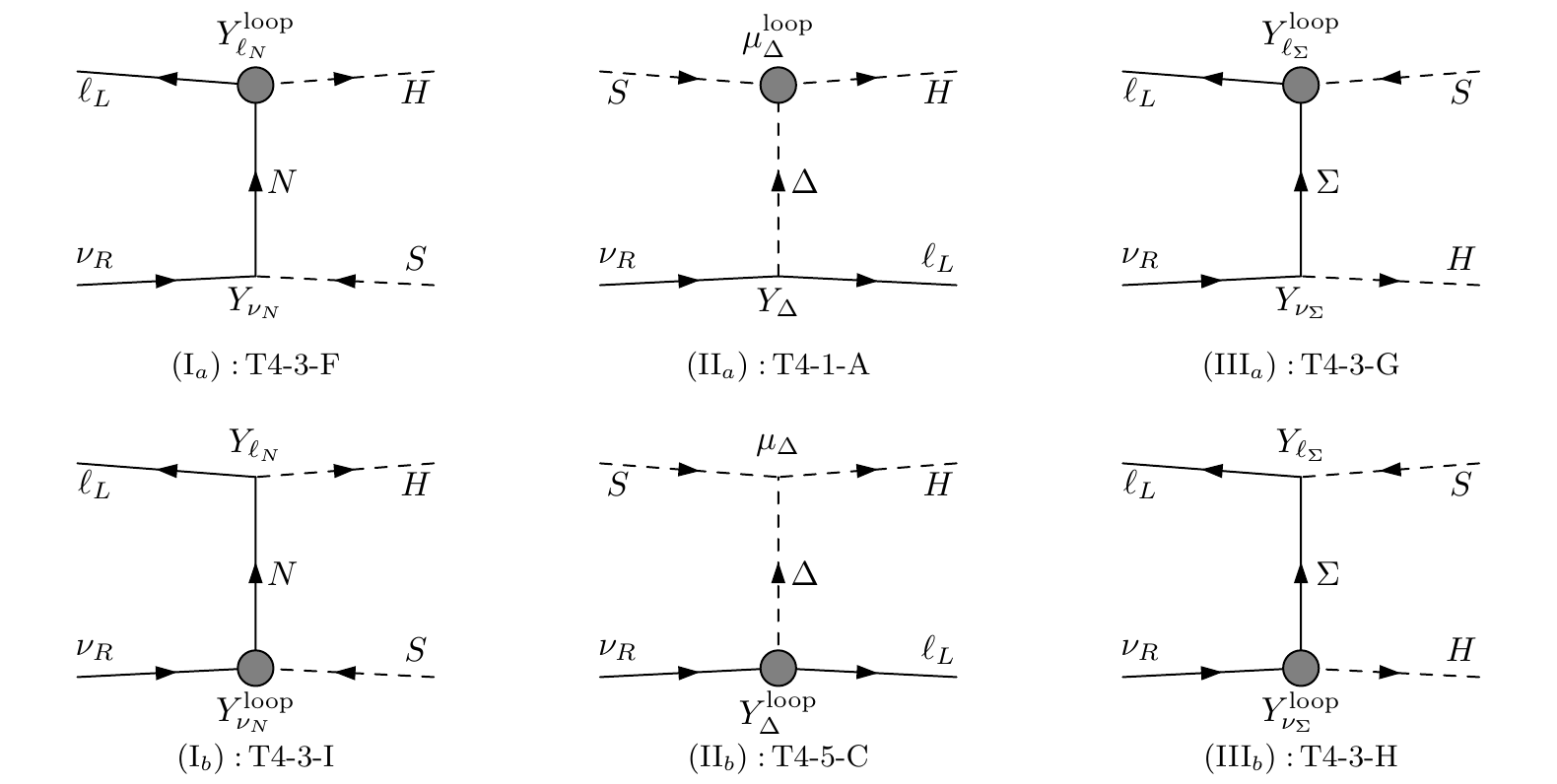}
\caption{\label{fig:seesaw_eff}Type-I (left), type-II (center) and type-III (right) Dirac seesaw realizations via one-loop vertices.}
\end{figure}

From table~\ref{tab:figT4}, we can see that all the diagrams based on topology T4 except for T4-3-I are just some minor corrections to the neutrino mass matrix, since the particle necessary for a tree level Dirac seesaw always exists and the tree level contributions can not be forbidden if no additional symmetry is imposed. Actually, the topology T4 can be interpreted as extensions of the tree level Dirac seesaw mechanisms at one-loop. These extensions are clearly illustrated in figure~\ref{fig:seesaw_eff}, where one vertex of the Dirac seesaw diagram is generated at one-loop level, we mark the effective one-loop vertex by a gray filled circle. On the other hand, for each Dirac seesaw mechanism, either vertex can be subject to one-loop corrections which exactly lead to one-loop diagrams based on topology T4. As a result, all the Feynman diagrams generated from T4 except T4-3-I can be discarded if one doesn't introduce additional discrete (or gauge) symmetries. The messenger particle $X_{1}$ could be a $SU(2)_L$ singlet, doublet or triplet as shown in tables~\ref{tab:figT4}. If $X_1$ is a singlet or doublet, it can be identified as $N$ and $\Sigma$ ($\Delta$ for $X_{1}$ being scalar) in the Dirac seesaw models respectively. As a consequence, the diagrams of topology T4 with singlet or doublet $X_{1}$ messenger always allow for the appearance of tree level Dirac seesaw. An exception is the diagram T4-3-I, we see that the new field $X_{1}$ is a $SU(2)_L$ triplet fermion for the solutions IV and V such that it can not be used as the mediator of the Dirac seesaw. Notice that the parameter $\alpha$ can't be equal to $\pm1$ for the solution IV otherwise one of the particles $X_{2}, X_{3}$ and $X_{4}$ can be identified as $\Delta\sim \mathbf{2}_{1}^{-}$ or $\Sigma\sim
\mathbf{2}_{-1}^{-}$ in the Dirac seesaw. Moreover, we can find out all genuine models generated from the diagram T4-3-I. $X_1$ should be a triplet fermion to forbid tree level contribution, the $SU(2)_{L}$ gauge invariance of the interaction vertex involving $S$ and $\nu_{R}$ entails that $X_{2}, X_{3}$
and $X_{4}$ have to transform in the same way under $SU(2)_{L}$. That is to say, $X_{2}, X_{3}$ and $X_{4}$ are all $SU(2)_{L}$ $n-$multiplet with $n\geq2$. Taking into account $U(1)_{Y}$ symmetry further, we find that the four mediators should transform as $X^{F}_1\sim\mathbf{3}^{+}_{0}$, $X^{F}_2\sim n^{\pm}_{\alpha}$,
$X^{S}_3\sim n^{\mp}_{\alpha}$ and $X^{S}_4\sim n^{\pm}_{\alpha}$, where $\alpha\neq\pm1$ for $n=2$.

It is remarkable that the models for radiative Dirac neutrino masses can be easily read off from tables~\ref{tab:figT1}--\ref{tab:figT4}.
For illustration purposes, we take T1-1-A-I as an example. The particle content of this model consists of three SM singlets $X_{1}^{F}\sim\mathbf{1}^{\mp}_{\alpha}$,
$X_{2}^{S}\sim\mathbf{1}^{\pm}_{\alpha}$, $X_{3}^{S}\sim\mathbf{1}^{\mp}_{\alpha}$, and one doublet
$X_{4}^{S}\sim\mathbf{2}^{\mp}_{\alpha-1}$. The relevant Feynman diagram can be generated from the following Lagrangian:
\begin{equation}
\begin{split}
\mathscr{L}_{\text{T1-1-A-I}}= &\left[a_{\alpha i}
\overline{\ell_{L\alpha}}X_{1,i}^c X_{4}+b_{i\beta}
\overline{X_{1,i}^c}\nu_{R\beta}X_{2}^{*}+
cX_{2}SX_{3}^{*}+dX_{3}H^{\dagger}\widetilde{X_{4}}+\text{H.c.}\right]\\
&-M_{X_{1}}\overline{X_{1}}X_{1}-M_{X_{2}}^{2}X_{2}^{\dagger}X_{2}-M_{X_{3}}^{2}X_{3}^{\dagger}X_{3}
-M_{X_{4}}^{2}X_{4}^{\dagger}X_{4}\,,
\end{split}
\end{equation}
where $\alpha, \beta, i=1, 2, 3$ are flavor indices, and we assume
there are three generations of the new fermion $X_{1}^{F}$. The parameters $a$ and $b$ denote the Yukawa couplings of the new interaction vertices, and $c, d$ are the coupling constants among the Higgs field $H$, singlet scalar field $S$ and the new scalars $X_{2}^{S}, X_{3}^{S}$ and $X_{4}^{S}$.

In order to obtain the effective mass operator, it is convenient to calculate the one-loop Feynman diagram of T1-1-A-I before the electroweak symmetry breaking. The neutrino masses are generated in case that the Higgs field $H$ and the singlet scalar $S$ acquire nonzero vacuum expectation values. Since the momenta of the external lines ($\ell_L$, $\nu_{R}$, $H$ and $S$) are irrelevant to the neutrino masses, we can set them to be zero for simplicity, and then the loop integral would be simplified considerably. After straightforward algebra, we find the neutrino mass matrix is given by
\begin{equation}
(m_{\nu})_{\alpha\beta}/(\langle H\rangle\langle S\rangle)=
-M_{X_{1}}^{(i)}a_{\alpha  i} b_{i \beta } c dI_{4}\left(M_{X_{2}},M_{X_{3}},M_{X_{4}},M_{X_{1}}^{(i)}\right)\,,
\end{equation}
where the loop function $I_4$ is given by Eq.~\eqref{eq:I4}.
One will encounter five different integrals $I_{2,3,4}$ and $J_{3,4}$
in the calculation of the one-loop models, their explicit
forms are showed in Appendix~\ref{sec:app_mass}. For all other
possible one-loop neutrino mass models, we can follow the same procedure to write out the Lagrange of the model and extract the resulting predictions
for neutrino masses.

\section{\label{sec:forbidtree}Forbidding lower order contributions in non-genuine models}

If the one-loop diagram for the dimension five operator
$\overline{\ell_{L\alpha}}\widetilde{H}\nu_{R\beta}S$ gives the leading order contribution to Dirac neutrino masses, the tree level Dirac seesaws have to be forbidden. In this section, we shall show that absence of lower order contributions can be achieved by exploiting non-abelian discrete flavor symmetry. As discussed above, almost all diagram generated from the topology T4 can be regarded as the one-loop extension of the
Dirac seesaw. For example, the T4-1-A-I model is a one-loop extension of the type-II Dirac seesaws. We shall demonstrate that the one-loop contribution to neutrino masses can be the dominant contribution and none of the particles necessary for tree level seesaws exists if the flavor symmetry $A_4$ is properly implemented in the T4-1-A-I model.

The $A_4$ group is simplest finite group with three dimensional irreducible representation. It has been widely exploited to predict lepton mixing angles and CP violation phases~\cite{Ding:2011gt,Ding:2013bpa,Li:2016nap,Ma:2017moj}. $A_4$
is the symmetry group of a regular tetrahedron, and it has 12 elements. The $A_4$ group can be generated by two generators $S$ and $T$ obeying the relations:
\begin{equation}
S^2=T^3=(ST)^3=1\,.
\end{equation}
$A_4$ has four inequivalent irreducible representations: three
singlets $\mathbf{1}$, $\mathbf{1}'$, $\mathbf{1}''$ and
one triplet $\mathbf{3}$. The multiplication rules of $A_4$ are as follows:
\begin{eqnarray}
\nonumber&&\mathbf{1}\otimes R=R,~~\mathbf{1'}\otimes\mathbf{1''}=\mathbf{1},~~~\mathbf{1'}\otimes\mathbf{1'}=\mathbf{1''},~~~\mathbf{1''}\otimes\mathbf{1''}=\mathbf{1'},\\
&&\mathbf{3}\otimes\mathbf{3}=\mathbf{1}\oplus\mathbf{1'}\oplus\mathbf{1''}\oplus\mathbf{3}\oplus\mathbf{3},~~\mathbf{3}\otimes\mathbf{1'}=\mathbf{3},~~\mathbf{3}\otimes\mathbf{1''}=\mathbf{3}\,,
\end{eqnarray}
where $R$ denotes any $A_4$ representation.
\begin{table}[t!]
\centering
\begin{tabular}{|c|c|c|c|c|c|c|c|c|}
\hline\hline
& \multicolumn{8}{c|}{T4-1-A-I} \\ \hline
\text{Fields} & $\ell_L$ & $\nu_R$ & $H$ & $S$ & $X_{1}^{S}$ & $X_{2}^{S}$ & $X_{3}^{S}$ & $X_{4}^{S}$ \\ \hline
\text{$Z_{2}$ \& Gauge Sym.} & $\mathbf{2}_{-1}^{+}$ & $\mathbf{1}_{0}^{-}$ & $\mathbf{2}_{1}^{+}$ & $\mathbf{1}_{0}^{-}$ & $\mathbf{2}_{1}^{-}$ & $\mathbf{1}_{\alpha}^{\pm}$ & $\mathbf{1}_{\alpha}^{\mp}$ & $\mathbf{2}_{\alpha-1}^{\mp}$ \\ \hline
\text{$A_4$ Sym.} & $\mathbf{3}$ & $\mathbf{3}$ & $\mathbf{1}$ & $\mathbf{1}$ & $\mathbf{3}$ & $\mathbf{3}$ & $\mathbf{3}$ & $\mathbf{3}$ \\ \hline\hline
\end{tabular}
\caption{\label{tab:qnumber_A4} Forbidding tree level contributions in the T4-1-A-I model by using the flavor symmetry $A_4$. Here we list the transformation properties of the fields under $Z_{2}$, $A_4$ and SM gauge symmetry.}
\end{table}

For the model T4-1-A-I, we assign the three generations of left-handed lepton doublets $\ell_{L}$ and right-handed neutrinos $\nu_{R}$ to two $A_4$ triplets $\mathbf{3}$, the Higgs doublet $H$ and singlet scalar $S$ are invariant under $A_4$. We assume that each messenger particle in T4-1-A-I has three families and it transforms as triplet. The transformation properties of all the fields under $A_4$ are collected in table~\ref{tab:qnumber_A4}. Obviously the $Z_{2}$ symmetry prevents the renormalizable Yukawa coupling $\overline{\ell_{L}}\widetilde{H}\nu_{R}$. Although the SM gauge
symmetry allows for $X_{1}^{S}\sim\mathbf{2}_{1}^{-1}$ to be identified as mediator particle of the type-II Dirac seesaw, the $A_{4}$ symmetry forbids the vertex $H^{\dagger}X_{1}S$. As a consequence, there is no tree level contribution to neutrino masses, and the non-vanishing neutrino masses dominantly arise from one-loop diagrams.

Usually one has more than one option to choose a flavor symmetry to forbid the tree level diagrams. In the above example, we use a $Z_{2}$ symmetry to forbid the direct Yukawa coupling, and use $A_{4}$ to forbid the appearance of type-II Dirac seesaw. In fact, both diagrams can be forbidden by a single flavor group $S_{4}$~\cite{Ma:2005pd,Ding:2009iy,Hagedorn:2010th,Ding:2013hpa}. $S_4$ is the permutation group of four distinct objects, and it is
isomorphic to the symmetry group of a regular octahedron. The $S_4$
group has five irreducible representations:  two singlets $\mathbf{1}$
and $\mathbf{1}^{\prime}$, one doublet $\mathbf{2}$ and two triplets
$\mathbf{3}$ and $\mathbf{3}^{\prime}$. The multiplication rules between various representations are given by
\begin{eqnarray}
\nonumber&&\mathbf{1^{\prime}}\otimes\mathbf{1^{\prime}}=\mathbf{1},\quad
\mathbf{1^{\prime}}\otimes\mathbf{2}=\mathbf{2},\quad
\mathbf{1^{\prime}}\otimes\mathbf{3}=\mathbf{3^{\prime}},\quad
\mathbf{1^{\prime}}\otimes\mathbf{3^{\prime}}=\mathbf{3},\\
&&
\mathbf{2}\otimes\mathbf{2}=\mathbf{1}\oplus\mathbf{1^{\prime}}\oplus\mathbf{2},\quad
\mathbf{2}\otimes\mathbf{3}=\mathbf{2}\otimes\mathbf{3^{\prime}}=\mathbf{3}\oplus\mathbf{3^{\prime}},\\
\nonumber&&\mathbf{3}\otimes\mathbf{3}=\mathbf{3^{\prime}}\otimes\mathbf{3^{\prime}}=\mathbf{1}\oplus\mathbf{2}\oplus\mathbf{3}\oplus\mathbf{3^{\prime}},\quad
\mathbf{3}\otimes\mathbf{3^{\prime}}=\mathbf{1^{\prime}}\oplus\mathbf{2}\oplus\mathbf{3}\oplus\mathbf{3^{\prime}}\,.
\end{eqnarray}
Here we can assign $\ell_{L}$ and $\nu_{R}$ to be $S_4$ triplets $\mathbf{3}$ and $\mathbf{3}^{\prime}$ respectively, the Higgs field $H$ and singlet scalar $S$ transform as $\mathbf{1}$, the new fields
$X_{2}^{S}, X_{3}^{S}$ and $X_{4}^{S}$ in the loop are assigned to
$S_{4}$ doublets, while $X_{1}^{S}$ transforms as a non-trivial singlet $\mathbf{1}'$. The transformation properties of all fields are displayed in table~\ref{tab:qnumber_S4}, we can see that the renormalizable Yukawa coupling is automatically forbidden by the $S_{4}$ symmetry, and the three-scalar interaction vertex $H^{\dagger}X_{1}S$ in type-II Dirac seesaw is also incompatible with $S_{4}$ symmetry. Therefore the one-loop diagram in T4-1-A-I model would be the leading order contribution to neutrino masses if the $S_4$ flavor symmetry is imposed in above way. In a similar fashion, discrete flavor symmetry can be used to forbid the tree level contribution of Dirac seesaw for other non-genuine models in table~\ref{tab:figT4}, and we expect it can also help to explain the values of the leptonic mixing angles and CP phases.

\begin{table}[hptb!]
\centering
\begin{tabular}{|c|c|c|c|c|c|c|c|c|}
\hline\hline
& \multicolumn{8}{c|}{T4-1-A-I} \\ \hline
\text{Fields} & $\ell_L$ & $\nu_R$ & $H$ & $S$ & $X_{1}^{S}$ & $X_{2}^{S}$ & $X_{3}^{S}$ & $X_{4}^{S}$ \\ \hline
\text{Gauge Sym.} & $\mathbf{2}_{-1}$ & $\mathbf{1}_{0}$ & $\mathbf{2}_{1}$ & $\mathbf{1}_{0}$ & $\mathbf{2}_{1}$ & $\mathbf{1}_{\alpha}$ & $\mathbf{1}_{\alpha}$ & $\mathbf{2}_{\alpha-1}$ \\ \hline
\text{$S_4$ Sym.} & $\mathbf{3}$ & $\mathbf{3^\prime}$ & $\mathbf{1}$ & $\mathbf{1}$ & $\mathbf{1}^{\prime}$ & $\mathbf{2}$ & $\mathbf{2}$ & $\mathbf{2}$ \\ \hline\hline
\end{tabular}
\caption{\label{tab:qnumber_S4} Forbidding tree level contributions in the T4-1-A-I model by using the $S_4$ flavor symmetry. Here we list the transformation properties of the fields under $S_4$ and SM gauge symmetry.}
\end{table}

\section{\label{sec:darkmatter} Dark matter candidates}

Neutrino masses and dark matter are the only experimental evidences for physics beyond standard model at present. The existence of non-baryonic dark matter is well established by using cosmological and astrophysical probes. However, dark matter has not been observed yet at direct detection experiments, indirect detection experiments or colliders. The exact nature of dark matter remains elusive. The most widely accepted hypothesis is that the dark matter is composed of weakly interacting massive particles, WIMPs, that interact only through gravity and the weak force. We expect that the underlying new physics model should be able to explain both neutrino masses and dark matter. It has been suggested that the neutrino masses and cosmological dark matter may be closely related to each other. For example, dark matter can be connected with the neutrino sector through the generation of the light neutrino masses~\cite{Ma:2006km}. The dark matter particle can play the role of messenger of radiative neutrino mass generation~\cite{Hirsch:2013ola}. So far many neutrino mass models containing dark matter particle have been proposed~\cite{Gu:2007gy,Gu:2007ug,Farzan:2012sa,Bonilla:2016diq,Chulia:2016ngi,Kanemura:2017haa}.
Some systematic analysis of radiative neutrino mass models with viable dark matter candidates has been performed in~\cite{Restrepo:2013aga,Yao:2017vtm,Simoes:2017kqb}.

Similar to the scotogenic model~\cite{Ma:2006km} for Majorana neutrinos, the particles mediating the one-loop diagram for Dirac neutrino mass generation could be dark matter candidates, such that both puzzles of neutrino masses and the dark matter can be explained in a same model. In the following, we shall discuss the possible viable dark matter candidates in our radiative neutrino mass models.

If dark matter is an elementary particle, the present experimental data dictate that it should be  stable, color and electrically neutral. In order to ensure the stability of dark matter particle, usually an extra discrete symmetry is added to distinguish dark matter particles from SM ones. This discrete symmetry is typically chosen to be $Z_{2}$ symmetry, which is widely used in radiative neutrino mass models to account for dark matter. Consequently a dark matter $Z'_2$ symmetry is imposed in our models as well. Notice that $Z'_2$ is distinct from the $Z_2$ symmetry which forbids the renormalizable Yukawa coupling. We assume that the new fields mediating the one-loop diagram for neutrino masses are odd while the remaining fields are even under $Z'_{2}$. For instance, for the diagram of topology T4, the fields $X_{2},X_{3}$ and $X_{4}$ transform as odd under $Z_{2}^{\prime}$ symmetry, $X_{1}$ as well as the SM particles are even under $Z_{2}^{\prime}$. The lightest odd particle among $X_{2},X_{3}$ and $X_{4}$ could be dark matter candidate if it is neutral.

It is well-known that the electric charge of a field is given by $Q=T_{3}+Y/2$, where $T_{3}$ is the isospin and $Y$ is the hypercharge. Requiring that dark matter is electrically neutral, we arrive at the constraint $Y=-2T_{3}$ which the dark matter candidate must satisfy. Moreover, dark matter direct detection experiments play a crucial role in our analysis. A electroweak multiplet containing a dark matter candidate has tree level interactions with quarks via $Z$ boson exchange, consequently the direct detection cross section via nucleon recoil is proportional to $Y^{2}$. If the hypercharge $Y$ is nonzero, scattering cross sections is generally quite large such that dark matter would have been observed by the current experiments PandaX~\cite{Fu:2016ega} and XENON1T~\cite{Aprile:2017iyp}. Combining with the electrically neutral condition, we find the dark matter candidate must satisfy $Y=T_{3}=0$. This requirement eliminates multiplets with even number of fields (i.e., doublets or quartets etc.). But the scalar doublet with hypercharge $Y=\pm1$ is an exception. In such models, a mass splitting can be enforced between the scalar and pseudo-scalar, and this can eliminate the coupling with the $Z$ boson at tree level. Thus the direct detection rate of the neutral component of the doublet (the dark matter candidate) is still lower than experimental bounds~\cite{LopezHonorez:2006gr,Farzan:2012ev}. We can obtain the values of the parameter $\alpha$ by solving the condition $Y=T_{3}=0$. Each such value of $\alpha$ determines a model which can simultaneously account for dark matter and neutrino masses,

In the following we shall take T3-1-A-I as an example to illustrate under what conditions the radiative neutrino mass models can also accommodate dark matter. This model contains three new fields: a singlet fermion $X_{1}^{F}\sim \mathbf{1}_{\alpha}^{\mp}$, a singlet scalar $X_{2}^{S}\sim \mathbf{1}_{\alpha}^{\pm}$, and a doublet scalar $X_{3}^{S}\sim \mathbf{2}_{\alpha-1}^{\mp}$. There are two possible values of $\alpha$ which are consistent with dark matter.
\begin{itemize}
\item $\alpha=0$: \quad $X_{1}^{0}$, \quad $X_{2}^{0}$, \quad
    $X_{3}^{S}=(X_{3}^{0},X_{3}^{-})$

In this case the dark matter candidate is a singlet fermion
or a mixture of the neutral components from the singlet and doublet scalars.
The fermion field $X_{1}^{0}$ should be vector-like to ensure the anomaly
cancellation.

\item $\alpha=2$: \quad $X_{1}^{+}$, \quad $X_{2}^{+}$, \quad
    $X_{3}^{S}=(X_{3}^{+},X_{3}^{0})$

Only $X_{3}^{S}$ has a neutral component, and $X_{3}^{0}$ is a viable dark matter candidate since it is a scalar doublet with $Y=1$. The anomaly cancellation entails that the fermion $X_{1}^{+}$ should be vector-like.

\end{itemize}

In the same manner the possible dark matter candidates can be discussed for the other viable one-loop models showed in tables~\ref{tab:figT1}--\ref{tab:figT4}, the corresponding
results are given in the next to the last column, where we display the values of $\alpha$ compatible with dark matter as subscripts outside the square bracket. We see that in most cases the $\alpha$ values compatible with dark matter are excluded by the requirement that the contributions from the tree level Dirac seesaw should be absent in a given model. However, after the dark matter $Z_{2}^{\prime}$ symmetry is taken into account, many values of $\alpha$ excluded by the tree level neutrino masses can survive, and these numbers are shadowed in grey. The reason is that the radiative neutrino mass messengers are odd under $Z'_2$ and they can not mediate the Dirac seesaw diagrams.

It is notable that some of these models contain doubly-charged fermions or scalars. One can expect interesting signatures for them at colliders, since the SM background is highly suppressed and the signals can be more easily searched for in the LHC data. The possibility of detecting such a particle at colliders has been discussed in the literature~\cite{Babu:2002uu,AristizabalSierra:2006gb,Nebot:2007bc,Alloul:2013raa,Babu:2016rcr},
and search for scalar doublets with exotic charges has been performed at  CMS~\cite{Chatrchyan:2012ya} and ATLAS~\cite{ATLAS:2012hi}.  The final collider signatures and the corresponding branching ratios would allow to distinguish between singlets, doublets and triplets, and fermions or scalars.

\section{\label{sec:conclusion}Conclusion}

In this work, we have systematically investigated the possible ultraviolet completions of a dimension five Dirac neutrino mass operator $\overline{\ell_{L\alpha}}\widetilde{H}\nu_{R\beta}S$ at both tree and one-loop levels. Similar to the famous seesaw mechanism for Majorana neutrinos, we find there are only three types of Dirac seesaw models which are able to generate Dirac neutrino masses at tree level. We give a systematic analysis of neutrino mass models at one-loop order and construct all possible one-loop topologies. We find that the radiative Dirac neutrino mass models can be classified into three categories: (A)~Models with divergent diagrams, the tree level contribution is necessary and the divergence can be absorbed by the counter terms to obtain finite neutrino masses. (B)~Models with non-genuine but finite diagrams, they can be regarded as extensions of the Dirac seesaw mechanisms  where one of the vertices is generated at one loop. In this case, one can introduce discrete flavor symmetry such as $A_4$ and $S_4$  to forbid the tree level Dirac seesaw diagrams. (C)~Models with genuine and finite diagrams, the one-loop diagram naturally gives the leading contribution to neutrino mass matrix, the tree level contribution is absent due to the structure and particle content of the model. We find there are in total eight diagrams T1-1-A, T1-1-B, T1-2-A,  T1-2-B, T1-3-D, T1-3-E, T3-1-A and T4-3-I which can lead to genuine one-loop Dirac neutrino mass models. We have listed the new mediator fields and their transformation rules under the SM gauge symmetry and $Z_{2}$ symmetry, where we consider scalars and fermions as mediators transforming as singlets, doublets or triplets of $SU(2)_{L}$. Moreover, we have presented the predictions for the neutrino mass matrix for each possible model. All these results are collected in tables~\ref{tab:figT1}--\ref{tab:figT4}.

The internal messenger in the one-loop diagram for Dirac neutrino mass generation could be dark matter candidate. The stability of the dark matter candidate is ensured by the addition of a $Z'_2$ symmetry.
A dark matter candidate should be electrically neutral, and should be consistent with current bounds from dark matter direct detection experiments. As a consequence, the dark matter could be either
singlet or triplet fermions and scalars with zero hypercharge, or a
scalar doublet with $Y=\pm1$ in the concerned radiative neutrino mass models. The possible dark matter candidates in each model are
presented in tables~\ref{tab:figT1}--\ref{tab:figT4}. Our results can be consistently and readily used to construct one-loop Dirac neutrino mass models which can also address dark matter issue.

These radiative neutrino mass models can be tested using different searches, in particular, we highlight these models containing new particles with exotic electric charges since they give rise to background-free signals which are amenable to searches at the LHC. It is very appealing to study the collider, dark matter and neutrino phenomenology of many of these viable models in detail. Since each radiative model has its own particularities, the phenomenological implications should be studied on a case-by-case basis and the possible signals would be very model dependent. Hence we leave the detailed analysis of the rich phenomenology of these models for future work.

\section*{Acknowledgements}
This work is supported by the National Natural Science Foundation of China under Grant No.11522546.

\clearpage

\begin{appendix}

\section{\label{sec:app_mass}Loop integrals }

In this appendix we explicitly give the integrals appearing in the neutrino masses when evaluating the one-loop diagrams.
\begin{eqnarray}
\nonumber I_{2} (M_{A}, M_{B})&\equiv & \int \frac{{\rm d}^{d} k}{(2\pi)^{d}i}\frac{1}{(k^{2} - M_{A}^{2})(k^{2} - M_{B}^{2})}\\
&=&\frac{1}{(4\pi)^{2}} \left[\frac{2}{\epsilon} -\gamma_E +1+\ln(4\pi)-\ln
M_B^2+\frac{M_A^2}{M_A^2-M_B^2}\ln\frac{M_B^2}{M_A^2}\right]\,.
\end{eqnarray}
\begin{eqnarray}
\nonumber I_{3} (M_{A}, M_{B}, M_{C})&\equiv & \int \frac{{\rm d}^{d}
k}{(2\pi)^{d} i}\frac{1}{(k^{2} - M_{A}^{2})(k^{2} - M_{B}^{2})(k^{2}
- M_{C}^{2})} \\
\nonumber&=&\frac{1}{(4\pi)^{2}}\left[\frac{M_{A}^{2}}{(M_{A}^{2} - M_{B}^{2})(M_{A}^{2}-M_{C}^{2})}\ln
\frac{M_{C}^{2}}{M_{A}^{2}}\right.\\
&&\hskip0.1in\left.+\frac{M_{B}^{2}}{(M_{B}^{2} -
M_{A}^{2})(M_{B}^{2} - M_{C}^{2})}\ln\frac{M_{C}^{2}}{M_{B}^{2}}\right]\,.
\end{eqnarray}
\begin{eqnarray}
\nonumber I_{4} (M_{A}, M_{B}, M_{C}, M_{D})&\equiv & \int \frac{{\rm
d}^{d} k}{(2\pi)^{d} i}\frac{1}{(k^{2} - M_{A}^{2})(k^{2} -
M_{B}^{2})(k^{2} - M_{C}^{2})(k^{2} - M_{D}^{2})} \\
\nonumber &= & \frac{1}{(4\pi)^{2}}\left[\frac{M_{A}^{2}}{(M_{A}^{2} - M_{B}^{2})(M_{A}^{2} -
M_{C}^{2})(M_{A}^{2} - M_{D}^{2})}\ln\frac{M_{D}^{2}}{M_{A}^{2}}\right.\\
\nonumber&& \hskip0.1in
+\frac{M_{B}^{2}}{(M_{B}^{2} -
M_{A}^{2})(M_{B}^{2} - M_{C}^{2})(M_{B}^{2} - M_{D}^{2})}\ln\frac{M_{D}^{2}}{M_{B}^{2}}\\
\label{eq:I4}&& \hskip0.1in
+\left.\frac{M_{C}^{2}}{(M_{C}^{2} -
M_{A}^{2})(M_{C}^{2} - M_{B}^{2})(M_{C}^{2} - M_{D}^{2})}\ln\frac{M_{D}^{2}}{M_{C}^{2}}\right]\,.
\end{eqnarray}
\begin{eqnarray}
\nonumber J_{3} (M_{A}, M_{B}, M_{C})&\equiv&\int \frac{{\rm d}^{d}
k}{(2\pi)^{d}i}\frac{k^{2}}{(k^{2} - M_{A}^{2})(k^{2} -
M_{B}^{2})(k^{2} - M_{C}^{2})}\\
\nonumber&=&\frac{1}{(4\pi)^{2}}\left[ \frac{2}{\epsilon}-\gamma_{E}+1+\ln(4\pi)-\ln
M_{C}^{2}\right. \\
&&\hskip-0.3in\left.+\frac{M_A^4}{(M_A^2-M_B^2)(M_A^2-M_C^2)}\ln\frac{M_C^2}{M_{A}^{2}}+
\frac{M_B^4}{(M_B^2-M_A^2)(M_B^2-M_C^2)}\ln\frac{M_{C}^{2}}{M_B^2}\right]\,.
\end{eqnarray}
\begin{eqnarray}
\nonumber J_{4} (M_{A}, M_{B}, M_{C}, M_{D})&\equiv&\int \frac{{\rm
d}^{d} k}{(2\pi)^{d} i}\frac{k^{2}}{(k^{2} - M_{A}^{2})(k^{2} -
M_{B}^{2})(k^{2} - M_{C}^{2})(k^{2} - M_{D}^{2})} \\
\nonumber&=&\frac{1}{(4\pi)^{2}}\left[\frac{M_{A}^{4}}{(M_{A}^{2} - M_{B}^{2})(M_{A}^{2} -
M_{C}^{2})(M_{A}^{2} - M_{D}^{2})}\ln\frac{M_{D}^{2}}{M_{A}^{2}}\right.\\
\nonumber&&\hskip0.15in
+\frac{M_{B}^{4}}{(M_{B}^{2} -
M_{A}^{2})(M_{B}^{2} - M_{C}^{2})(M_{B}^{2} - M_{D}^{2})}\ln \frac{M_{D}^{2}}{M_{B}^{2}}\\
&&\hskip0.15in
\left.+\frac{M_{C}^{4}}{(M_{C}^{2} -
M_{A}^{2})(M_{C}^{2} - M_{B}^{2})(M_{C}^{2} - M_{D}^{2})}\ln\frac{M_{D}^{2}}{M_{C}^{2}}\right]\,,
\end{eqnarray}
where $\epsilon=4-d$ is an infinitely small quantity, and $\gamma_E$ is the Euler-Mascheroni constant, we can see that the functions $I_2$ and $J_3$ are divergent, and the other functions are finite.

\section{\label{sec:app_table}Field assignments, neutrino masses and dark matter candidates}

As explained in section~\ref{sec:treeloop}, the finite one-loop diagrams are generated from the topologies T1, T3 and T4 shown in figure~\ref{fig:topfigloopA}. In this appendix, we shall present the possible quantum numbers of the messenger fields and the expressions of the neutrino mass matrix for each diagram in tables~\ref{tab:figT1}--\ref{tab:figT4}, where the messengers are assumed to transform as singlets, doublets or triplets under $SU(2)_L$. If the one-loop diagrams give the leading contributions to neutrino masses, the diagrams of Dirac seesaws should be absent so that certain values of $\alpha$ are excluded. In addition, we list the possible dark matter candidates and the corresponding values of $\alpha$ which are shown as subscript outside the square bracket. The neutral component of the particles mediating the loop diagram, either a fermion or a scalar, is a good dark matter candidate. The $Z_{2}^{\prime}$ symmetry stabilizing dark matter can also help to forbid the tree level diagrams. As a consequence, many values of $\alpha$ excluded by the disappearance of tree level diagrams become admissible after the $Z_{2}^{\prime}$ symmetry is taken into account, and all these numbers are shadowed in grey in tables~\ref{tab:figT1}--\ref{tab:figT4}.
Hence we conclude that both neutrino masses and dark matter could be explained simultaneously in these viable one-loop models.

\begin{table}[hptb!]
\small{
\centering
\begin{tabular}{|c|c|c|c|c|c|c|c|c|c|}
\hline\hline
\multirow{2}{*}{Topology} & \multirow{2}{*}{Sol.} & \multirow{2}{*}{$X_1^F$} & \multirow{2}{*}{$X_2^S$} & \multirow{2}{*}{$X_3^S$} & \multirow{2}{*}{$X_4^S$} & \multicolumn{2}{c|}{Excluded $\alpha$} & \multirow{2}{*}{Dark Matter} & Exotic\\ \cline{7-8}
 &  &  &  &  &  & $Z_2^{I}$ & $Z_2^{II}$ & & charges\\ \hline
\multirow{6}{*}{T1-1-A} & I & $\mathbf{1}_{\alpha}^{\mp}$ & $\mathbf{1}_{\alpha}^{\pm}$ & $\mathbf{1}_{\alpha}^{\mp}$ & $\mathbf{2}_{\alpha -1}^{\mp}$ & \hl{$0, 2$} & \hl{$0$} & $[X_1, X_2, X_3, X_4]_{0}$, $[X_4]_{2}$ & \ding{55} \\ \cline{2-10}
 & II & $\mathbf{2}_{\alpha}^{\mp}$ & $\mathbf{2}_{\alpha}^{\pm}$ & $\mathbf{2}_{\alpha}^{\mp}$ & $\mathbf{1}_{\alpha -1}^{\mp}$ & \hl{$\pm 1$} & \hl{$\pm 1$} & $[X_2, X_3]_{-1}$, $[X_2, X_3, X_4]_{1}$ & \ding{55} \\ \cline{2-10}
 & \multirow{2}{*}{III} & \multirow{2}{*}{$\mathbf{2}_{\alpha}^{\mp}$} & \multirow{2}{*}{$\mathbf{2}_{\alpha}^{\pm}$} & \multirow{2}{*}{$\mathbf{2}_{\alpha}^{\mp}$} & \multirow{2}{*}{$\mathbf{3}_{\alpha -1}^{\mp}$} & \multirow{2}{*}{\hl{$\pm 1$}} & \multirow{2}{*}{\hl{$\pm 1$}} & $[X_2, X_3, X_4]_{-1}$ & \ding{51} \\ \cline{9-10}
 & & & & & & & & $[X_2, X_3, X_4]_{1}$ & \ding{55} \\ \cline{2-10}
 & \multirow{2}{*}{IV} & \multirow{2}{*}{$\mathbf{3}_{\alpha}^{\mp}$} & \multirow{2}{*}{$\mathbf{3}_{\alpha}^{\pm}$} & \multirow{2}{*}{$\mathbf{3}_{\alpha}^{\mp}$} & \multirow{2}{*}{$\mathbf{2}_{\alpha -1}^{\mp}$} & \multirow{2}{*}{\hl{$0, 2$}} & \multirow{2}{*}{$\varnothing$} & $[X_1, X_2, X_3, X_4]_{0}$ & \ding{55} \\ \cline{9-10}
 & & & & & & & & $[X_2, X_3, X_4]_{2}$ & \ding{51} \\ \hline
\multirow{6}{*}{T1-1-B} & I & $\mathbf{1}_{\alpha}^{\mp}$ & $\mathbf{1}_{\alpha}^{\pm}$ & $\mathbf{2}_{\alpha -1}^{\pm}$ & $\mathbf{2}_{\alpha -1}^{\mp}$ & \hl{$0, 2$} & \hl{$0, 2$} & $[X_1, X_2, X_3, X_4]_{0}$, $[X_3, X_4]_{2}$ & \ding{55} \\ \cline{2-10}
 & II & $\mathbf{2}_{\alpha}^{\mp}$ & $\mathbf{2}_{\alpha}^{\pm}$ & $\mathbf{1}_{\alpha -1}^{\pm}$ & $\mathbf{1}_{\alpha -1}^{\mp}$ & \hl{$\pm 1$} & \hl{$\pm 1$} & $[X_2]_{-1}$, $[X_2, X_3, X_4]_{1}$ & \ding{55} \\ \cline{2-10}
 & \multirow{2}{*}{III} & \multirow{2}{*}{$\mathbf{2}_{\alpha}^{\mp}$} & \multirow{2}{*}{$\mathbf{2}_{\alpha}^{\pm}$} & \multirow{2}{*}{$\mathbf{3}_{\alpha -1}^{\pm}$} & \multirow{2}{*}{$\mathbf{3}_{\alpha -1}^{\mp}$} & \multirow{2}{*}{\hl{$\pm 1$}} & \multirow{2}{*}{\hl{$\pm 1$}} & $[X_2, X_3, X_4]_{-1}$ & \ding{51} \\ \cline{9-10}
 & & & & & & & & $[X_2, X_3, X_4]_{1}$ & \ding{55} \\ \cline{2-10}
 & \multirow{2}{*}{IV} & \multirow{2}{*}{$\mathbf{3}_{\alpha}^{\mp}$} & \multirow{2}{*}{$\mathbf{3}_{\alpha}^{\pm}$} & \multirow{2}{*}{$\mathbf{2}_{\alpha -1}^{\pm}$} & \multirow{2}{*}{$\mathbf{2}_{\alpha -1}^{\mp}$} & \multirow{2}{*}{\hl{$0, 2$}} & \multirow{2}{*}{\hl{$0, 2$}} & $[X_1, X_2, X_3, X_4]_{0}$ & \ding{55} \\ \cline{9-10}
 & & & & & & & & $[X_2, X_3, X_4]_{2}$ & \ding{51} \\ \hline
\multicolumn{10}{|m{0.9\linewidth}<{\centering}|}{$(m_{\nu})_{\alpha\beta}/(\langle H\rangle\langle S\rangle)=-M_{X_{1}}^{(i)}a_{\alpha  i} b_{i \beta } c dI_{4}\left(M_{X_{2}},M_{X_{3}},M_{X_{4}},M_{X_{1}}^{(i)}\right)$}\\ \hline\hline
\multirow{2}{*}{Topology} & \multirow{2}{*}{Sol.} & \multirow{2}{*}{$X_1^S$} & \multirow{2}{*}{$X_2^F$} & \multirow{2}{*}{$X_3^F$} & \multirow{2}{*}{$X_4^F$} & \multicolumn{2}{c|}{Excluded $\alpha$} & \multirow{2}{*}{Dark Matter} & Exotic\\ \cline{7-8}
 &  &  &  &  &  & $Z_2^{I}$ & $Z_2^{II}$ & & charges \\ \hline
\multirow{5}{*}{T1-2-A} & I & $\mathbf{1}_{\alpha}^{\mp}$ & $\mathbf{1}_{\alpha}^{\pm}$ & $\mathbf{1}_{\alpha}^{\mp}$ & $\mathbf{2}_{\alpha -1}^{\mp}$ & \hl{$0, 2$} & \hl{$0$} & $[X_1, X_2, X_3, X_4]_{0}$ & \ding{55} \\ \cline{2-10}
 & II & $\mathbf{2}_{\alpha}^{\mp}$ & $\mathbf{2}_{\alpha}^{\pm}$ & $\mathbf{2}_{\alpha}^{\mp}$ & $\mathbf{1}_{\alpha -1}^{\mp}$ & \hl{$\pm 1$} & \hl{$\pm 1$} & $[X_1]_{-1}$, $[X_1, X_2, X_3, X_4]_{1}$ & \ding{55} \\ \cline{2-10}
 & \multirow{2}{*}{III} & \multirow{2}{*}{$\mathbf{2}_{\alpha}^{\mp}$} & \multirow{2}{*}{$\mathbf{2}_{\alpha}^{\pm}$} & \multirow{2}{*}{$\mathbf{2}_{\alpha}^{\mp}$} & \multirow{2}{*}{$\mathbf{3}_{\alpha -1}^{\mp}$} & \multirow{2}{*}{\hl{$\pm 1$}} & \multirow{2}{*}{\hl{$\pm 1$}} & $[X_1]_{-1}$ & \ding{51} \\ \cline{9-10}
 & & & & & & & & $[X_1, X_2, X_3, X_4]_{1}$ & \ding{55} \\ \cline{2-10}
 & IV & $\mathbf{3}_{\alpha}^{\mp}$ & $\mathbf{3}_{\alpha}^{\pm}$ & $\mathbf{3}_{\alpha}^{\mp}$ & $\mathbf{2}_{\alpha -1}^{\mp}$ & \hl{$0, 2$} & $\varnothing$ & $[X_1, X_2, X_3, X_4]_{0}$ & \ding{55} \\ \hline
\multirow{5}{*}{T1-2-B} & I & $\mathbf{1}_{\alpha}^{\mp}$ & $\mathbf{1}_{\alpha}^{\pm}$ & $\mathbf{2}_{\alpha -1}^{\pm}$ & $\mathbf{2}_{\alpha -1}^{\mp}$ & \hl{$0, 2$} & \hl{$0, 2$} & $[X_1, X_2, X_3, X_4]_{0}$ & \ding{55} \\ \cline{2-10}
 & II & $\mathbf{2}_{\alpha}^{\mp}$ & $\mathbf{2}_{\alpha}^{\pm}$ & $\mathbf{1}_{\alpha -1}^{\pm}$ & $\mathbf{1}_{\alpha -1}^{\mp}$ & \hl{$\pm 1$} & \hl{$\pm 1$} & $[X_1]_{-1}$, $[X_1, X_2, X_3, X_4]_{1}$ & \ding{55} \\ \cline{2-10}
 & \multirow{2}{*}{III} & \multirow{2}{*}{$\mathbf{2}_{\alpha}^{\mp}$} & \multirow{2}{*}{$\mathbf{2}_{\alpha}^{\pm}$} & \multirow{2}{*}{$\mathbf{3}_{\alpha -1}^{\pm}$} & \multirow{2}{*}{$\mathbf{3}_{\alpha -1}^{\mp}$} & \multirow{2}{*}{\hl{$\pm 1$}} & \multirow{2}{*}{\hl{$\pm 1$}} & $[X_1]_{-1}$ & \ding{51} \\ \cline{9-10}
 & & & & & & & & $[X_1, X_2, X_3, X_4]_{1}$ & \ding{55} \\ \cline{2-10}
 & IV & $\mathbf{3}_{\alpha}^{\mp}$ & $\mathbf{3}_{\alpha}^{\pm}$ & $\mathbf{2}_{\alpha -1}^{\pm}$ & $\mathbf{2}_{\alpha -1}^{\mp}$ & \hl{$0, 2$} & \hl{$0, 2$} & $[X_1, X_2, X_3, X_4]_{0}$ & \ding{55} \\ \hline
\multicolumn{10}{|m{0.9\linewidth}<{\centering}|}{$(m_{\nu})_{\alpha\beta}/(\langle H\rangle\langle S\rangle)=-a_{\alpha  i} d_{i j} c_{j k} b_{k \beta }\left[M_{X_{4}}^{(i)} M_{X_{3}}^{(j)} M_{X_{2}}^{(k)} I_{4}\left(M_{X_{1}},M_{X_{2}}^{(k)},M_{X_{3}}^{(j)},M_{X_{4}}^{(i)}\right)\right.$ $\left.+\left(M_{X_{4}}^{(i)}+M_{X_{3}}^{(j)}+M_{X_{2}}^{(k)}\right) J_{4}\left(M_{X_{1}},M_{X_{2}}^{(k)},M_{X_{3}}^{(j)},M_{X_{4}}^{(i)}\right)\right]$}\\ \hline\hline
\multirow{2}{*}{Topology} & \multirow{2}{*}{Sol.} & \multirow{2}{*}{$X_1^F$} & \multirow{2}{*}{$X_2^S$} & \multirow{2}{*}{$X_3^S$} & \multirow{2}{*}{$X_4^F$} & \multicolumn{2}{c|}{Excluded $\alpha$} & \multirow{2}{*}{Dark Matter} & Exotic \\ \cline{7-8}
 &  &  &  &  &  & $Z_2^{I}$ & $Z_2^{II}$ & & charges \\ \hline
\multirow{5}{*}{T1-3-D} & I & $\mathbf{1}_{\alpha}^{\mp}$ & $\mathbf{1}_{\alpha}^{\pm}$ & $\mathbf{1}_{\alpha}^{\mp}$ & $\mathbf{2}_{\alpha +1}^{\mp}$ & \hl{$-2, 0$} & \hl{$0$} & $[X_1, X_2, X_3, X_4]_{0}$ & \ding{55} \\ \cline{2-10}
 & II & $\mathbf{2}_{\alpha}^{\mp}$ & $\mathbf{2}_{\alpha}^{\pm}$ & $\mathbf{2}_{\alpha}^{\mp}$ & $\mathbf{1}_{\alpha +1}^{\mp}$ & \hl{$\pm 1$} & \hl{$\pm 1$} & $[X_1, X_2, X_3, X_4]_{-1}$, $[X_2, X_3]_{1}$ & \ding{55} \\ \cline{2-10}
 & \multirow{2}{*}{III} & \multirow{2}{*}{$\mathbf{2}_{\alpha}^{\mp}$} & \multirow{2}{*}{$\mathbf{2}_{\alpha}^{\pm}$} & \multirow{2}{*}{$\mathbf{2}_{\alpha}^{\mp}$} & \multirow{2}{*}{$\mathbf{3}_{\alpha +1}^{\mp}$} & \multirow{2}{*}{\hl{$\pm 1$}} & \multirow{2}{*}{\hl{$\pm 1$}} & $[X_1, X_2, X_3, X_4]_{-1}$ & \ding{55} \\ \cline{9-10}
 & & & & & & & & $[X_2, X_3]_{1}$ & \ding{51} \\ \cline{2-10}
 & IV & $\mathbf{3}_{\alpha}^{\mp}$ & $\mathbf{3}_{\alpha}^{\pm}$ & $\mathbf{3}_{\alpha}^{\mp}$ & $\mathbf{2}_{\alpha +1}^{\mp}$ & \hl{$-2, 0$} & $\varnothing$ & $[X_1, X_2, X_3, X_4]_{0}$ & \ding{55} \\ \hline
\multirow{6}{*}{T1-3-E} & I & $\mathbf{1}_{\alpha}^{\mp}$ & $\mathbf{1}_{\alpha}^{\pm}$ & $\mathbf{2}_{\alpha -1}^{\pm}$ & $\mathbf{1}_{\alpha}^{\pm}$ & \hl{$0$} & \hl{$0, 2$} & $[X_1, X_2, X_3, X_4]_{0}$, $[X_3]_{2}$ & \ding{55} \\ \cline{2-10}
 & II & $\mathbf{2}_{\alpha}^{\mp}$ & $\mathbf{2}_{\alpha}^{\pm}$ & $\mathbf{1}_{\alpha -1}^{\pm}$ & $\mathbf{2}_{\alpha}^{\pm}$ & \hl{$\pm 1$} & \hl{$\pm 1$} & $[X_2]_{-1}$, $[X_2, X_3]_{1}$ & \ding{55} \\ \cline{2-10}
 & \multirow{2}{*}{III} & \multirow{2}{*}{$\mathbf{2}_{\alpha}^{\mp}$} & \multirow{2}{*}{$\mathbf{2}_{\alpha}^{\pm}$} & \multirow{2}{*}{$\mathbf{3}_{\alpha -1}^{\pm}$} & \multirow{2}{*}{$\mathbf{2}_{\alpha}^{\pm}$} & \multirow{2}{*}{\hl{$\pm 1$}} & \multirow{2}{*}{\hl{$\pm 1$}} & $[X_2, X_3]_{-1}$ & \ding{51} \\ \cline{9-10}
 & & & & & & & & $[X_2, X_3]_{1}$ & \ding{55} \\ \cline{2-10}
 & \multirow{2}{*}{IV} & \multirow{2}{*}{$\mathbf{3}_{\alpha}^{\mp}$} & \multirow{2}{*}{$\mathbf{3}_{\alpha}^{\pm}$} & \multirow{2}{*}{$\mathbf{2}_{\alpha -1}^{\pm}$} & \multirow{2}{*}{$\mathbf{3}_{\alpha}^{\pm}$} & \multirow{2}{*}{$\varnothing$} & \multirow{2}{*}{\hl{$0, 2$}} & $[X_1, X_2, X_3, X_4]_{0}$ & \ding{55} \\ \cline{9-10}
 & & & & & & & & $[X_2, X_3]_{2}$ & \ding{51} \\ \hline
\multicolumn{10}{|m{0.9\linewidth}<{\centering}|}{$(m_{\nu})_{\alpha\beta}/(\langle H\rangle\langle S\rangle)=-d_{\alpha  i} a_{i j} b_{j \beta } c\left[M_{X_{4}}^{(i)} M_{X_{1}}^{(j)} I_{4}\left(M_{X_{2}},M_{X_{3}},M_{X_{1}}^{(j)},M_{X_{4}}^{(i)}\right)\right.$ $\left.+J_{4}\left(M_{X_{2}},M_{X_{3}},M_{X_{1}}^{(j)},M_{X_{4}}^{(i)}\right)\right]$}\\ \hline\hline
\end{tabular}
\caption{\label{tab:figT1} The finite one-loop diagrams generated from the topology T1. We show the possible quantum numbers of the messenger fields, the predictions for neutrino masses, and the dark matter candidates. The absence of tree level Dirac seesaw excludes certain values of $\alpha$, where $\varnothing$ and $\mathbb{U}$ denote empty set and universal set respectively. The dark matter $Z'_2$ symmetry can prevent tree level contributions to neutrino masses, such that the excluded $\alpha$ values become admissible and they are shadowed in grey. }}
\end{table}

\begin{table}[hptb!]
\centering
\begin{tabular}{|c|c|c|c|c|c|c|c|c|}
\hline\hline
\multirow{2}{*}{Topology} & \multirow{2}{*}{Sol.} & \multirow{2}{*}{$X_1^F$} & \multirow{2}{*}{$X_2^S$} & \multirow{2}{*}{$X_3^S$} & \multicolumn{2}{c|}{Excluded $\alpha$} & \multirow{2}{*}{Dark Matter} & Exotic \\ \cline{6-7}
 &  &  &  &  & $Z_2^{I}$ & $Z_2^{II}$ & & charges \\ \hline
\multirow{6}{*}{T3-1-A} & I & $\mathbf{1}_{\alpha}^{\mp}$ & $\mathbf{1}_{\alpha}^{\pm}$ & $\mathbf{2}_{\alpha -1}^{\mp}$ & \hl{$0, 2$} & \hl{$0$} & $[X_1, X_2, X_3]_{0}$, $[X_3]_{2}$ & \ding{55} \\ \cline{2-9}
 & II & $\mathbf{2}_{\alpha}^{\mp}$ & $\mathbf{2}_{\alpha}^{\pm}$ & $\mathbf{1}_{\alpha -1}^{\mp}$ & \hl{$\pm 1$} & \hl{$\pm 1$} & $[X_2]_{-1}$, $[X_2, X_3]_{1}$ & \ding{55} \\ \cline{2-9}
 & \multirow{2}{*}{III} & \multirow{2}{*}{$\mathbf{2}_{\alpha}^{\mp}$} & \multirow{2}{*}{$\mathbf{2}_{\alpha}^{\pm}$} & \multirow{2}{*}{$\mathbf{3}_{\alpha -1}^{\mp}$} & \multirow{2}{*}{\hl{$\pm 1$}} & \multirow{2}{*}{\hl{$\pm 1$}} & $[X_2, X_3]_{-1}$ & \ding{51} \\ \cline{8-9}
 & & & & & & & $[X_2, X_3]_{1}$ & \ding{55} \\ \cline{2-9}
 & \multirow{2}{*}{IV} & \multirow{2}{*}{$\mathbf{3}_{\alpha}^{\mp}$} & \multirow{2}{*}{$\mathbf{3}_{\alpha}^{\pm}$} & \multirow{2}{*}{$\mathbf{2}_{\alpha -1}^{\mp}$} & \multirow{2}{*}{\hl{$0, 2$}} & \multirow{2}{*}{$\varnothing$} & $[X_1, X_2, X_3]_{0}$ & \ding{55} \\ \cline{8-9}
 & & & & & & & $[X_2, X_3]_{2}$ & \ding{51} \\ \hline
\multicolumn{9}{|m{0.75\linewidth}<{\centering}|}{$(m_{\nu})_{\alpha\beta}/(\langle H\rangle\langle S\rangle)=M_{X_{1}}^{(i)}a_{\alpha  i} b_{i \beta } cI_{3}\left(M_{X_{2}},M_{X_{3}},M_{X_{1}}^{(i)}\right)$}\\ \hline\hline
\end{tabular}
\caption{\label{tab:figT3} The finite one-loop diagrams generated from the topology T3. We show the possible quantum numbers of the messenger fields, the predictions for neutrino masses, and the dark matter candidates. The absence of tree level Dirac seesaw excludes certain values of $\alpha$, where $\varnothing$ and $\mathbb{U}$ denote empty set and universal set respectively. The dark matter $Z'_2$ symmetry can prevent tree level contributions to neutrino masses, such that the excluded $\alpha$ values become admissible and they are shadowed in grey.}
\end{table}

\begin{table}[hptb!]
\renewcommand\arraystretch{0.87}
\small{
\centering
\begin{tabular}{|c|c|c|c|c|c|c|c|c|c|}
\hline\hline
\multirow{2}{*}{Topology} & \multirow{2}{*}{Sol.} & \multirow{2}{*}{$X_1^S$} & \multirow{2}{*}{$X_2^S$} & \multirow{2}{*}{$X_3^S$} & \multirow{2}{*}{$X_4^S$} & \multicolumn{2}{c|}{Excluded $\alpha$} & \multirow{2}{*}{Dark Matter} & Exotic \\ \cline{7-8}
 &  &  &  &  &  & $Z_2^{I}$ & $Z_2^{II}$ & & charges \\ \hline
\multirow{6}{*}{T4-1-A} & I & $\mathbf{2}_{1}^{-}$ & $\mathbf{1}_{\alpha}^{\pm}$ & $\mathbf{1}_{\alpha}^{\mp}$ & $\mathbf{2}_{\alpha -1}^{\mp}$ & $\mathbb{U}$ & $\mathbb{U}$ & $[X_2, X_3, X_4]_{0}$, $[X_4]_{2}$ & \ding{55} \\ \cline{2-10}
 & II & $\mathbf{2}_{1}^{-}$ & $\mathbf{2}_{\alpha}^{\pm}$ & $\mathbf{2}_{\alpha}^{\mp}$ & $\mathbf{1}_{\alpha -1}^{\mp}$ & $\mathbb{U}$ & $\mathbb{U}$ & $[X_2, X_3]_{-1}$, $[X_2, X_3, X_4]_{1}$ & \ding{55} \\ \cline{2-10}
 & \multirow{2}{*}{III} & \multirow{2}{*}{$\mathbf{2}_{1}^{-}$} & \multirow{2}{*}{$\mathbf{2}_{\alpha}^{\pm}$} & \multirow{2}{*}{$\mathbf{2}_{\alpha}^{\mp}$} & \multirow{2}{*}{$\mathbf{3}_{\alpha -1}^{\mp}$} & \multirow{2}{*}{$\mathbb{U}$} & \multirow{2}{*}{$\mathbb{U}$} & $[X_2, X_3, X_4]_{-1}$ & \ding{51} \\ \cline{9-10}
 & & & & & & & & $[X_2, X_3, X_4]_{1}$ & \ding{55} \\ \cline{2-10}
 & \multirow{2}{*}{IV} & \multirow{2}{*}{$\mathbf{2}_{1}^{-}$} & \multirow{2}{*}{$\mathbf{3}_{\alpha}^{\pm}$} & \multirow{2}{*}{$\mathbf{3}_{\alpha}^{\mp}$} & \multirow{2}{*}{$\mathbf{2}_{\alpha -1}^{\mp}$} & \multirow{2}{*}{$\mathbb{U}$} &  \multirow{2}{*}{ $\mathbb{U}$} & $[X_2, X_3, X_4]_{0}$ & \ding{55} \\ \cline{9-10}
 & & & & & & & & $[X_2, X_3, X_4]_{2}$ & \ding{51} \\ \hline
\multicolumn{10}{|m{0.8\linewidth}<{\centering}|}{$(m_{\nu})_{\alpha\beta}/(\langle H\rangle\langle S\rangle)=\frac{a_{\alpha  \beta } b c d}{M_{X_{1}}^2}I_{3}\left(M_{X_{2}},M_{X_{3}},M_{X_{4}}\right)$}\\ \hline\hline
\multirow{2}{*}{Topology} & \multirow{2}{*}{Sol.} & \multirow{2}{*}{$X_1^F$} & \multirow{2}{*}{$X_2^F$} & \multirow{2}{*}{$X_3^S$} & \multirow{2}{*}{$X_4^S$} & \multicolumn{2}{c|}{Excluded $\alpha$} & \multirow{2}{*}{Dark Matter} & Exotic \\ \cline{7-8}
 &  &  &  &  &  & $Z_2^{I}$ & $Z_2^{II}$ & & charges \\ \hline
\multirow{6}{*}{T4-3-F} & I & $\mathbf{1}_{0}^{+}$ & $\mathbf{1}_{\alpha}^{\pm}$ & $\mathbf{2}_{\alpha +1}^{\pm}$ & $\mathbf{1}_{\alpha}^{\pm}$ & $\mathbb{U}$ & $\mathbb{U}$ & $[X_3]_{-2}$, $[X_2, X_3, X_4]_{0}$ & \ding{55} \\ \cline{2-10}
 & II & $\mathbf{1}_{0}^{+}$ & $\mathbf{2}_{\alpha}^{\pm}$ & $\mathbf{1}_{\alpha +1}^{\pm}$ & $\mathbf{2}_{\alpha}^{\pm}$ & $\mathbb{U}$ & $\mathbb{U}$ & $[X_3, X_4]_{-1}$, $[X_4]_{1}$ & \ding{55} \\ \cline{2-10}
 & \multirow{2}{*}{III} & \multirow{2}{*}{$\mathbf{1}_{0}^{+}$} & \multirow{2}{*}{$\mathbf{2}_{\alpha}^{\pm}$} & \multirow{2}{*}{$\mathbf{3}_{\alpha +1}^{\pm}$} & \multirow{2}{*}{$\mathbf{2}_{\alpha}^{\pm}$} & \multirow{2}{*}{$\mathbb{U}$} & \multirow{2}{*}{$\mathbb{U}$} & $[X_3, X_4]_{-1}$ & \ding{55} \\ \cline{9-10}
 & & & & & & & & $[X_3, X_4]_{1}$ & \ding{51} \\ \cline{2-10}
 & \multirow{2}{*}{IV} & \multirow{2}{*}{$\mathbf{1}_{0}^{+}$} & \multirow{2}{*}{$\mathbf{3}_{\alpha}^{\pm}$} & \multirow{2}{*}{$\mathbf{2}_{\alpha +1}^{\pm}$} & \multirow{2}{*}{$\mathbf{3}_{\alpha}^{\pm}$} & \multirow{2}{*}{$\mathbb{U}$} & \multirow{2}{*}{$\mathbb{U}$} & $[X_3, X_4]_{-2}$ & \ding{51} \\ \cline{9-10}
 & & & & & & & & $[X_2, X_3, X_4]_{0}$ & \ding{55} \\ \hline
\multirow{5}{*}{T4-3-G} & I & $\mathbf{2}_{-1}^{-}$ & $\mathbf{1}_{\alpha}^{\pm}$ & $\mathbf{2}_{\alpha +1}^{\pm}$ & $\mathbf{2}_{\alpha +1}^{\mp}$ & $\mathbb{U}$ & $\mathbb{U}$ & $[X_3, X_4]_{-2}$, $[X_2, X_3, X_4]_{0}$ & \ding{55} \\ \cline{2-10}
 & II & $\mathbf{2}_{-1}^{-}$ & $\mathbf{2}_{\alpha}^{\pm}$ & $\mathbf{1}_{\alpha +1}^{\pm}$ & $\mathbf{1}_{\alpha +1}^{\mp}$ & $\mathbb{U}$ & $\mathbb{U}$ & $[X_3, X_4]_{-1}$ & \ding{55} \\ \cline{2-10}
 & III & $\mathbf{2}_{-1}^{-}$ & $\mathbf{2}_{\alpha}^{\pm}$ & $\mathbf{3}_{\alpha +1}^{\pm}$ & $\mathbf{3}_{\alpha +1}^{\mp}$ & $\mathbb{U}$ & $\mathbb{U}$ & $[X_3, X_4]_{-1}$ & \ding{55} \\ \cline{2-10}
 & \multirow{2}{*}{IV} & \multirow{2}{*}{$\mathbf{2}_{-1}^{-}$} & \multirow{2}{*}{$\mathbf{3}_{\alpha}^{\pm}$} & \multirow{2}{*}{$\mathbf{2}_{\alpha +1}^{\pm}$} & \multirow{2}{*}{$\mathbf{2}_{\alpha +1}^{\mp}$} & \multirow{2}{*}{$\mathbb{U}$} & \multirow{2}{*}{$\mathbb{U}$} & $[X_3, X_4]_{-2}$ & \ding{51} \\ \cline{9-10}
 & & & & & & & & $[X_2, X_3, X_4]_{0}$ & \ding{55} \\ \hline
\multicolumn{10}{|m{0.8\linewidth}<{\centering}|}{$(m_{\nu})_{\alpha\beta}/(\langle H\rangle\langle S\rangle)=\frac{M_{X_{2}}^{(i)}}{M_{X_{1}}^{(j)}}c_{\alpha  i} b_{i j} a_{j \beta } dI_{3}\left(M_{X_{3}},M_{X_{4}},M_{X_{2}}^{(i)}\right)$}\\ \hline\hline
\multirow{2}{*}{Topology} & \multirow{2}{*}{Sol.} & \multirow{2}{*}{$X_1^F$} & \multirow{2}{*}{$X_2^F$} & \multirow{2}{*}{$X_3^S$} & \multirow{2}{*}{$X_4^S$} & \multicolumn{2}{c|}{Excluded $\alpha$} & \multirow{2}{*}{Dark Matter} & Exotic \\ \cline{7-8}
 &  &  &  &  &  & $Z_2^{I}$ & $Z_2^{II}$ & & charges \\ \hline
\multirow{6}{*}{T4-3-H} & I & $\mathbf{2}_{1}^{-}$ & $\mathbf{1}_{\alpha}^{\pm}$ & $\mathbf{1}_{\alpha}^{\mp}$ & $\mathbf{2}_{\alpha -1}^{\mp}$ & $\mathbb{U}$ & $\mathbb{U}$ & $[X_2, X_3, X_4]_{0}$, $[X_4]_{2}$ & \ding{55} \\ \cline{2-10}
 & II & $\mathbf{2}_{1}^{-}$ & $\mathbf{2}_{\alpha}^{\pm}$ & $\mathbf{2}_{\alpha}^{\mp}$ & $\mathbf{1}_{\alpha -1}^{\mp}$ & $\mathbb{U}$ & $\mathbb{U}$ & $[X_3]_{-1}$, $[X_3, X_4]_{1}$ & \ding{55} \\ \cline{2-10}
 & \multirow{2}{*}{III} & \multirow{2}{*}{$\mathbf{2}_{1}^{-}$} & \multirow{2}{*}{$\mathbf{2}_{\alpha}^{\pm}$} & \multirow{2}{*}{$\mathbf{2}_{\alpha}^{\mp}$} & \multirow{2}{*}{$\mathbf{3}_{\alpha -1}^{\mp}$} & \multirow{2}{*}{$\mathbb{U}$} & \multirow{2}{*}{$\mathbb{U}$} & $[X_3, X_4]_{-1}$ & \ding{51} \\ \cline{9-10}
 & & & & & & & & $[X_3, X_4]_{1}$ & \ding{55} \\ \cline{2-10}
 & \multirow{2}{*}{IV} & \multirow{2}{*}{$\mathbf{2}_{1}^{-}$} & \multirow{2}{*}{$\mathbf{3}_{\alpha}^{\pm}$} & \multirow{2}{*}{$\mathbf{3}_{\alpha}^{\mp}$} & \multirow{2}{*}{$\mathbf{2}_{\alpha -1}^{\mp}$} & \multirow{2}{*}{$\mathbb{U}$} & \multirow{2}{*}{$\mathbb{U}$} & $[X_2, X_3, X_4]_{0}$ & \ding{55} \\ \cline{9-10}
 & & & & & & & & $[X_3, X_4]_{2}$ & \ding{51} \\ \hline
\multirow{5}{*}{T4-3-I} & I & $\mathbf{1}_{0}^{+}$ & $\mathbf{1}_{\alpha}^{\pm}$ & $\mathbf{1}_{\alpha}^{\mp}$ & $\mathbf{1}_{\alpha}^{\pm}$ & $\mathbb{U}$ & $\mathbb{U}$ & $[X_2, X_3, X_4]_{0}$ & \ding{55} \\ \cline{2-10}
 & II & $\mathbf{1}_{0}^{+}$ & $\mathbf{2}_{\alpha}^{\pm}$ & $\mathbf{2}_{\alpha}^{\mp}$ & $\mathbf{2}_{\alpha}^{\pm}$ & $\mathbb{U}$ & $\mathbb{U}$ & $[X_3, X_4]_{-1}$, $[X_3, X_4]_{1}$ & \ding{55} \\ \cline{2-10}
 & III & $\mathbf{1}_{0}^{+}$ & $\mathbf{3}_{\alpha}^{\pm}$ & $\mathbf{3}_{\alpha}^{\mp}$ & $\mathbf{3}_{\alpha}^{\pm}$ & $\mathbb{U}$ & $\mathbb{U}$ & $[X_2, X_3, X_4]_{0}$ & \ding{55} \\ \cline{2-10}
 & IV & $\mathbf{3}_{0}^{+}$ & $\mathbf{2}_{\alpha}^{\pm}$ & $\mathbf{2}_{\alpha}^{\mp}$ & $\mathbf{2}_{\alpha}^{\pm}$ & \hl{$\pm 1$} & \hl{$\pm 1$} & $[X_3, X_4]_{-1}$, $[X_3, X_4]_{1}$ & \ding{55} \\ \cline{2-10}
 & V & $\mathbf{3}_{0}^{+}$ & $\mathbf{3}_{\alpha}^{\pm}$ & $\mathbf{3}_{\alpha}^{\mp}$ & $\mathbf{3}_{\alpha}^{\pm}$ & $\varnothing$ & $\varnothing$ & $[X_2, X_3, X_4]_{0}$ & \ding{55} \\ \hline
\multicolumn{10}{|m{0.8\linewidth}<{\centering}|}{$(m_{\nu})_{\alpha\beta}/(\langle H\rangle\langle S\rangle)=\frac{M_{X_{2}}^{(j)}}{M_{X_{1}}^{(i)}}a_{\alpha  i} b_{i j} c_{j \beta } dI_{3}\left(M_{X_{3}},M_{X_{4}},M_{X_{2}}^{(j)}\right)$}\\ \hline\hline
\multirow{2}{*}{Topology} & \multirow{2}{*}{Sol.} & \multirow{2}{*}{$X_1^S$} & \multirow{2}{*}{$X_2^S$} & \multirow{2}{*}{$X_3^F$} & \multirow{2}{*}{$X_4^S$} & \multicolumn{2}{c|}{Excluded $\alpha$} & \multirow{2}{*}{Dark Matter} & Exotic \\ \cline{7-8}
 &  &  &  &  &  & $Z_2^{I}$ & $Z_2^{II}$ & & charges \\ \hline
\multirow{6}{*}{T4-5-C} & I & $\mathbf{2}_{-1}^{-}$ & $\mathbf{1}_{\alpha}^{\pm}$ & $\mathbf{1}_{\alpha}^{\mp}$ & $\mathbf{2}_{\alpha +1}^{\mp}$ & $\mathbb{U}$ & $\mathbb{U}$ & $[X_4]_{-2}$, $[X_2, X_3, X_4]_{0}$ & \ding{55} \\ \cline{2-10}
 & II & $\mathbf{2}_{-1}^{-}$ & $\mathbf{2}_{\alpha}^{\pm}$ & $\mathbf{2}_{\alpha}^{\mp}$ & $\mathbf{1}_{\alpha +1}^{\mp}$ & $\mathbb{U}$ & $\mathbb{U}$ & $[X_2, X_4]_{-1}$, $[X_2]_{1}$ & \ding{55} \\ \cline{2-10}
 & \multirow{2}{*}{III} & \multirow{2}{*}{$\mathbf{2}_{-1}^{-}$} & \multirow{2}{*}{$\mathbf{2}_{\alpha}^{\pm}$} & \multirow{2}{*}{$\mathbf{2}_{\alpha}^{\mp}$} & \multirow{2}{*}{$\mathbf{3}_{\alpha +1}^{\mp}$} & \multirow{2}{*}{$\mathbb{U}$} & \multirow{2}{*}{$\mathbb{U}$} & $[X_2, X_4]_{-1}$ & \ding{55} \\ \cline{9-10}
 & & & & & & & & $[X_2, X_4]_{1}$ & \ding{51} \\ \cline{2-10}
 & \multirow{2}{*}{IV} & \multirow{2}{*}{$\mathbf{2}_{-1}^{-}$} & \multirow{2}{*}{$\mathbf{3}_{\alpha}^{\pm}$} & \multirow{2}{*}{$\mathbf{3}_{\alpha}^{\mp}$} & \multirow{2}{*}{$\mathbf{2}_{\alpha +1}^{\mp}$} & \multirow{2}{*}{$\mathbb{U}$} & \multirow{2}{*}{$\mathbb{U}$} & $[X_2, X_4]_{-2}$ & \ding{51} \\ \cline{9-10}
 & & & & & & & & $[X_2, X_3, X_4]_{0}$ & \ding{55} \\ \hline
\multicolumn{10}{|m{0.8\linewidth}<{\centering}|}{$(m_{\nu})_{\alpha\beta}/(\langle H\rangle\langle S\rangle)=\frac{M_{X_{3}}^{(i)}}{M_{X_{1}}^2}d_{\alpha  i} c_{i \beta } a bI_{3}\left(M_{X_{2}},M_{X_{4}},M_{X_{3}}^{(i)}\right)$}\\ \hline\hline
\end{tabular}
\caption{\label{tab:figT4}The finite one-loop diagrams generated from the topology T4. We show the possible quantum numbers of the messenger fields, the predictions for neutrino masses, and the dark matter candidates. The absence of tree level Dirac seesaw excludes certain values of $\alpha$, where $\varnothing$ and $\mathbb{U}$ denote empty set and universal set respectively. The dark matter $Z'_2$ symmetry can prevent tree level contributions to neutrino masses, such that the excluded $\alpha$ values become admissible and they are shadowed in grey.}}
\end{table}

\section{\label{sec:within}Models contained within others}

When a model is contained within another model with more field multiplets, the neutrino masses predicted by the latter model receive two contributions, one from the topology of the model itself and another from the model it contains. When we calculate the neutrino mass matrix and confront with experimental data on neutrino masses and mixing angles, both contributions from these two diagrams should be taken into account. The inclusion relations of the viable models are indicated in table~\ref{tab:within}. This table shows the models with 3
multiplets which are contained in models with 4 multiplets.
\begin{table}[htbp!]
\centering
\begin{tabular}{|c|c||c|c|}
\hline\hline
\multirow{2}{*}{Model} & \multirow{2}{*}{$\alpha$}& \multicolumn{2}{c|}{Contained in}\\
\cline{3-4}
 & & Model & $\alpha$ \\
\hline
\multirow{4}{*}{T3-1-A-I}  & \multirow{2}{*}{$0$} & T1-1-A-I & $0$ \\ \cline{3-4}
  &  & T1-1-B-I & $0$ \\ \cline{2-4}
  & \multirow{2}{*}{$2$} & T1-1-A-I & $2$ \\ \cline{3-4}
  &  & T1-1-B-I & $2$ \\ \hline
\multirow{4}{*}{T3-1-A-II} & \multirow{2}{*}{$-1$} & T1-1-A-II & $-1$ \\ \cline{3-4}
  &  & T1-1-B-II & $-1$ \\ \cline{2-4}
  & \multirow{2}{*}{$1$} & T1-1-A-II & $1$ \\ \cline{3-4}
  &  & T1-1-B-II & $1$ \\ \hline
\multirow{4}{*}{T3-1-A-III} & \multirow{2}{*}{$-1$} & T1-1-A-III & $-1$ \\ \cline{3-4}
  &  & T1-1-B-III & $-1$ \\ \cline{2-4}
  & \multirow{2}{*}{$1$} & T1-1-A-III & $1$ \\ \cline{3-4}
  &  & T1-1-B-III & $1$ \\ \hline
\multirow{4}{*}{T3-1-A-IV}  & \multirow{2}{*}{$0$} & T1-1-A-IV & $0$ \\ \cline{3-4}
  &  & T1-1-B-IV & $0$ \\ \cline{2-4}
  & \multirow{2}{*}{$2$} & T1-1-A-IV & $2$ \\ \cline{3-4}
  &  & T1-1-B-IV & $2$ \\ \hline\hline
\end{tabular}
\caption{\label{tab:within} Models with 3 multiplets within models with 4 multiplets.}
\end{table}

\end{appendix}

\bibliographystyle{utphys}
\bibliography{numass}

\providecommand{\href}[2]{#2}\begingroup\raggedright\begin{thebibliography}{10}

\bibitem{Kajita:2016cak}
T.~Kajita, ``{Nobel Lecture: Discovery of atmospheric neutrino oscillations},''
\href{http://dx.doi.org/10.1103/RevModPhys.88.030501}{{\em Rev. Mod. Phys.}
  {\bfseries 88} no.~3, (2016) 030501}.

\bibitem{McDonald:2016ixn}
A.~B. McDonald, ``{Nobel Lecture: The Sudbury Neutrino Observatory: Observation
  of flavor change for solar neutrinos},''
\href{http://dx.doi.org/10.1103/RevModPhys.88.030502}{{\em Rev. Mod. Phys.}
  {\bfseries 88} no.~3, (2016) 030502}.

\bibitem{Bonnet:2012kz}
F.~Bonnet, M.~Hirsch, T.~Ota, and W.~Winter, ``{Systematic study of the d=5
  Weinberg operator at one-loop order},''
  \href{http://dx.doi.org/10.1007/JHEP07(2012)153}{{\em JHEP} {\bfseries 07}
  (2012) 153},
\href{http://arxiv.org/abs/1204.5862}{{\ttfamily arXiv:1204.5862 [hep-ph]}}.

\bibitem{Sierra:2014rxa}
D.~Aristizabal~Sierra, A.~Degee, L.~Dorame, and M.~Hirsch, ``{Systematic
  classification of two-loop realizations of the Weinberg operator},''
  \href{http://dx.doi.org/10.1007/JHEP03(2015)040}{{\em JHEP} {\bfseries 03}
  (2015) 040},
\href{http://arxiv.org/abs/1411.7038}{{\ttfamily arXiv:1411.7038 [hep-ph]}}.

\bibitem{Simoes:2017kqb}
C.~Simoes and D.~Wegman, ``{Radiative Two-Loop Neutrino Masses with Dark
  Matter},'' \href{http://dx.doi.org/10.1007/JHEP04(2017)148}{{\em JHEP}
  {\bfseries 04} (2017) 148},
\href{http://arxiv.org/abs/1702.04759}{{\ttfamily arXiv:1702.04759 [hep-ph]}}.

\bibitem{Cepedello:2017eqf}
R.~Cepedello, M.~Hirsch, and J.~C. Helo, ``{Loop neutrino masses from $d = 7$
  operator},'' \href{http://dx.doi.org/10.1007/JHEP07(2017)079}{{\em JHEP}
  {\bfseries 07} (2017) 079},
\href{http://arxiv.org/abs/1705.01489}{{\ttfamily arXiv:1705.01489 [hep-ph]}}.

\bibitem{Farzan:2012ev}
Y.~Farzan, S.~Pascoli, and M.~A. Schmidt, ``{Recipes and Ingredients for
  Neutrino Mass at Loop Level},''
  \href{http://dx.doi.org/10.1007/JHEP03(2013)107}{{\em JHEP} {\bfseries 03}
  (2013) 107},
\href{http://arxiv.org/abs/1208.2732}{{\ttfamily arXiv:1208.2732 [hep-ph]}}.

\bibitem{Minkowski:1977sc}
P.~Minkowski, ``{$\mu \to e\gamma$ at a Rate of One Out of $10^{9}$ Muon
  Decays?},''
\href{http://dx.doi.org/10.1016/0370-2693(77)90435-X}{{\em Phys. Lett.}
  {\bfseries B67} (1977) 421--428}.

\bibitem{Yanagida:1979as}
T.~Yanagida, ``{HORIZONTAL SYMMETRY AND MASSES OF NEUTRINOS},''
{\em Conf. Proc.} {\bfseries C7902131} (1979) 95--99.

\bibitem{GellMann:1980vs}
M.~Gell-Mann, P.~Ramond, and R.~Slansky, ``{Complex Spinors and Unified
  Theories},'' {\em Conf. Proc.} {\bfseries C790927} (1979) 315--321,
\href{http://arxiv.org/abs/1306.4669}{{\ttfamily arXiv:1306.4669 [hep-th]}}.

\bibitem{Mohapatra:1979ia}
R.~N. Mohapatra and G.~Senjanovic, ``{Neutrino Mass and Spontaneous Parity
  Violation},''
\href{http://dx.doi.org/10.1103/PhysRevLett.44.912}{{\em Phys. Rev. Lett.}
  {\bfseries 44} (1980) 912}.

\bibitem{Schechter:1980gr}
J.~Schechter and J.~W.~F. Valle, ``{Neutrino Masses in SU(2) x U(1)
  Theories},''
\href{http://dx.doi.org/10.1103/PhysRevD.22.2227}{{\em Phys. Rev.} {\bfseries
  D22} (1980) 2227}.

\bibitem{Roncadelli:1983ty}
M.~Roncadelli and D.~Wyler, ``{Naturally Light Dirac Neutrinos in Gauge
  Theories},''
\href{http://dx.doi.org/10.1016/0370-2693(83)90156-9}{{\em Phys. Lett.}
  {\bfseries 133B} (1983) 325--329}.

\bibitem{Ma:2015raa}
E.~Ma and R.~Srivastava, ``{Dirac or inverse seesaw neutrino masses from gauged
  $B–L$ symmetry},'' \href{http://dx.doi.org/10.1142/S0217732315300207}{{\em
  Mod. Phys. Lett.} {\bfseries A30} no.~26, (2015) 1530020},
\href{http://arxiv.org/abs/1504.00111}{{\ttfamily arXiv:1504.00111 [hep-ph]}}.

\bibitem{Chulia:2016ngi}
S.~Centelles~Chuliá, E.~Ma, R.~Srivastava, and J.~W.~F. Valle, ``{Dirac
  Neutrinos and Dark Matter Stability from Lepton Quarticity},''
  \href{http://dx.doi.org/10.1016/j.physletb.2017.01.070}{{\em Phys. Lett.}
  {\bfseries B767} (2017) 209--213},
\href{http://arxiv.org/abs/1606.04543}{{\ttfamily arXiv:1606.04543 [hep-ph]}}.

\bibitem{CentellesChulia:2017koy}
S.~Centelles~Chuliá, R.~Srivastava, and J.~W.~F. Valle, ``{Generalized
  Bottom-Tau unification, neutrino oscillations and dark matter: predictions
  from a lepton quarticity flavor approach},''
  \href{http://dx.doi.org/10.1016/j.physletb.2017.07.065}{{\em Phys. Lett.}
  {\bfseries B773} (2017) 26--33},
\href{http://arxiv.org/abs/1706.00210}{{\ttfamily arXiv:1706.00210 [hep-ph]}}.

\bibitem{Gu:2006dc}
P.-H. Gu and H.-J. He, ``{Neutrino Mass and Baryon Asymmetry from Dirac
  Seesaw},'' \href{http://dx.doi.org/10.1088/1475-7516/2006/12/010}{{\em JCAP}
  {\bfseries 0612} (2006) 010},
\href{http://arxiv.org/abs/hep-ph/0610275}{{\ttfamily arXiv:hep-ph/0610275
  [hep-ph]}}.

\bibitem{Valle:2016kyz}
J.~W.~F. Valle and C.~A. Vaquera-Araujo, ``{Dynamical seesaw mechanism for
  Dirac neutrinos},''
  \href{http://dx.doi.org/10.1016/j.physletb.2016.02.031}{{\em Phys. Lett.}
  {\bfseries B755} (2016) 363--366},
\href{http://arxiv.org/abs/1601.05237}{{\ttfamily arXiv:1601.05237 [hep-ph]}}.

\bibitem{Bonilla:2016zef}
C.~Bonilla and J.~W.~F. Valle, ``{Naturally light neutrinos in $Diracon$
  model},'' \href{http://dx.doi.org/10.1016/j.physletb.2016.09.022}{{\em Phys.
  Lett.} {\bfseries B762} (2016) 162--165},
\href{http://arxiv.org/abs/1605.08362}{{\ttfamily arXiv:1605.08362 [hep-ph]}}.

\bibitem{Bonilla:2017ekt}
C.~Bonilla, J.~M. Lamprea, E.~Peinado, and J.~W.~F. Valle, ``{Flavour-symmetric
  type-II Dirac neutrino seesaw mechanism},''
\href{http://arxiv.org/abs/1710.06498}{{\ttfamily arXiv:1710.06498 [hep-ph]}}.

\bibitem{Gu:2016hxh}
P.-H. Gu, ``{Peccei-Quinn symmetry for Dirac seesaw and leptogenesis},''
  \href{http://dx.doi.org/10.1088/1475-7516/2016/07/004}{{\em JCAP} {\bfseries
  1607} no.~07, (2016) 004},
\href{http://arxiv.org/abs/1603.05070}{{\ttfamily arXiv:1603.05070 [hep-ph]}}.

\bibitem{Mohapatra:1987nx}
R.~N. Mohapatra, ``{Left-right Symmetry and Finite One Loop Dirac Neutrino
  Mass},''
\href{http://dx.doi.org/10.1016/0370-2693(88)90610-7}{{\em Phys. Lett.}
  {\bfseries B201} (1988) 517--524}.

\bibitem{Gu:2007ug}
P.-H. Gu and U.~Sarkar, ``{Radiative Neutrino Mass, Dark Matter and
  Leptogenesis},'' \href{http://dx.doi.org/10.1103/PhysRevD.77.105031}{{\em
  Phys. Rev.} {\bfseries D77} (2008) 105031},
\href{http://arxiv.org/abs/0712.2933}{{\ttfamily arXiv:0712.2933 [hep-ph]}}.

\bibitem{Kanemura:2011jj}
S.~Kanemura, T.~Nabeshima, and H.~Sugiyama, ``{Neutrino Masses from
  Loop-Induced Dirac Yukawa Couplings},''
  \href{http://dx.doi.org/10.1016/j.physletb.2011.07.047}{{\em Phys. Lett.}
  {\bfseries B703} (2011) 66--70},
\href{http://arxiv.org/abs/1106.2480}{{\ttfamily arXiv:1106.2480 [hep-ph]}}.

\bibitem{Farzan:2012sa}
Y.~Farzan and E.~Ma, ``{Dirac neutrino mass generation from dark matter},''
  \href{http://dx.doi.org/10.1103/PhysRevD.86.033007}{{\em Phys. Rev.}
  {\bfseries D86} (2012) 033007},
\href{http://arxiv.org/abs/1204.4890}{{\ttfamily arXiv:1204.4890 [hep-ph]}}.

\bibitem{Ma:2016mwh}
E.~Ma and O.~Popov, ``{Pathways to Naturally Small Dirac Neutrino Masses},''
  \href{http://dx.doi.org/10.1016/j.physletb.2016.11.027}{{\em Phys. Lett.}
  {\bfseries B764} (2017) 142--144},
\href{http://arxiv.org/abs/1609.02538}{{\ttfamily arXiv:1609.02538 [hep-ph]}}.

\bibitem{Wang:2016lve}
W.~Wang and Z.-L. Han, ``{Naturally Small Dirac Neutrino Mass with Intermediate
  $SU(2)_{L}$ Multiplet Fields},''
  \href{http://dx.doi.org/10.1007/JHEP04(2017)166}{{\em JHEP} {\bfseries 04}
  (2017) 166},
\href{http://arxiv.org/abs/1611.03240}{{\ttfamily arXiv:1611.03240 [hep-ph]}}.

\bibitem{Ma:2017kgb}
E.~Ma and U.~Sarkar, ``{Radiative Left-Right Dirac Neutrino Mass},''
  \href{http://dx.doi.org/10.1016/j.physletb.2017.08.071}{{\em Phys. Lett.}
  {\bfseries B776} (2018) 54--57},
\href{http://arxiv.org/abs/1707.07698}{{\ttfamily arXiv:1707.07698 [hep-ph]}}.

\bibitem{Yao:2017vtm}
C.-Y. Yao and G.-J. Ding, ``{Systematic Study of One-Loop Dirac Neutrino Masses
  and Viable Dark Matter Candidates},''
  \href{http://dx.doi.org/10.1103/PhysRevD.96.095004}{{\em Phys. Rev.}
  {\bfseries D96} no.~9, (2017) 095004},
\href{http://arxiv.org/abs/1707.09786}{{\ttfamily arXiv:1707.09786 [hep-ph]}}.

\bibitem{Bonilla:2016diq}
C.~Bonilla, E.~Ma, E.~Peinado, and J.~W.~F. Valle, ``{Two-loop Dirac neutrino
  mass and WIMP dark matter},''
  \href{http://dx.doi.org/10.1016/j.physletb.2016.09.027}{{\em Phys. Lett.}
  {\bfseries B762} (2016) 214--218},
\href{http://arxiv.org/abs/1607.03931}{{\ttfamily arXiv:1607.03931 [hep-ph]}}.

\bibitem{Borah:2017leo}
D.~Borah and A.~Dasgupta, ``{Naturally Light Dirac Neutrino in Left-Right
  Symmetric Model},''
  \href{http://dx.doi.org/10.1088/1475-7516/2017/06/003}{{\em JCAP} {\bfseries
  1706} no.~06, (2017) 003},
\href{http://arxiv.org/abs/1702.02877}{{\ttfamily arXiv:1702.02877 [hep-ph]}}.

\bibitem{Wang:2017mcy}
W.~Wang, R.~Wang, Z.-L. Han, and J.-Z. Han, ``{The $B-L$ Scotogenic Models for
  Dirac Neutrino Masses},''
  \href{http://dx.doi.org/10.1140/epjc/s10052-017-5446-9}{{\em Eur. Phys. J.}
  {\bfseries C77} no.~12, (2017) 889},
\href{http://arxiv.org/abs/1705.00414}{{\ttfamily arXiv:1705.00414 [hep-ph]}}.

\bibitem{Hahn:2000kx}
T.~Hahn, ``{Generating Feynman diagrams and amplitudes with FeynArts 3},''
  \href{http://dx.doi.org/10.1016/S0010-4655(01)00290-9}{{\em Comput. Phys.
  Commun.} {\bfseries 140} (2001) 418--431},
\href{http://arxiv.org/abs/hep-ph/0012260}{{\ttfamily arXiv:hep-ph/0012260
  [hep-ph]}}.

\bibitem{Ding:2011gt}
G.-J. Ding and D.~Meloni, ``{A Model for Tri-bimaximal Mixing from a Completely
  Broken $A_4$},''
  \href{http://dx.doi.org/10.1016/j.nuclphysb.2011.10.001}{{\em Nucl. Phys.}
  {\bfseries B855} (2012) 21--45},
\href{http://arxiv.org/abs/1108.2733}{{\ttfamily arXiv:1108.2733 [hep-ph]}}.

\bibitem{Ding:2013bpa}
G.-J. Ding, S.~F. King, and A.~J. Stuart, ``{Generalised CP and $A_4$ Family
  Symmetry},'' \href{http://dx.doi.org/10.1007/JHEP12(2013)006}{{\em JHEP}
  {\bfseries 12} (2013) 006},
\href{http://arxiv.org/abs/1307.4212}{{\ttfamily arXiv:1307.4212 [hep-ph]}}.

\bibitem{Li:2016nap}
C.-C. Li, J.-N. Lu, and G.-J. Ding, ``{$A_4$ and CP symmetry and a model with
  maximal CP violation},''
  \href{http://dx.doi.org/10.1016/j.nuclphysb.2016.09.005}{{\em Nucl. Phys.}
  {\bfseries B913} (2016) 110--131},
\href{http://arxiv.org/abs/1608.01860}{{\ttfamily arXiv:1608.01860 [hep-ph]}}.

\bibitem{Ma:2017moj}
E.~Ma and G.~Rajasekaran, ``{Cobimaximal neutrino mixing from $A_4$ and its
  possible deviation},''
  \href{http://dx.doi.org/10.1209/0295-5075/119/31001}{{\em EPL} {\bfseries
  119} no.~3, (2017) 31001},
\href{http://arxiv.org/abs/1708.02208}{{\ttfamily arXiv:1708.02208 [hep-ph]}}.

\bibitem{Ma:2005pd}
E.~Ma, ``{Neutrino mass matrix from S(4) symmetry},''
  \href{http://dx.doi.org/10.1016/j.physletb.2005.10.019}{{\em Phys. Lett.}
  {\bfseries B632} (2006) 352--356},
\href{http://arxiv.org/abs/hep-ph/0508231}{{\ttfamily arXiv:hep-ph/0508231
  [hep-ph]}}.

\bibitem{Ding:2009iy}
G.-J. Ding, ``{Fermion Masses and Flavor Mixings in a Model with S(4) Flavor
  Symmetry},'' \href{http://dx.doi.org/10.1016/j.nuclphysb.2009.10.021}{{\em
  Nucl. Phys.} {\bfseries B827} (2010) 82--111},
\href{http://arxiv.org/abs/0909.2210}{{\ttfamily arXiv:0909.2210 [hep-ph]}}.

\bibitem{Hagedorn:2010th}
C.~Hagedorn, S.~F. King, and C.~Luhn, ``{A SUSY GUT of Flavour with S4 x SU(5)
  to NLO},'' \href{http://dx.doi.org/10.1007/JHEP06(2010)048}{{\em JHEP}
  {\bfseries 06} (2010) 048},
\href{http://arxiv.org/abs/1003.4249}{{\ttfamily arXiv:1003.4249 [hep-ph]}}.

\bibitem{Ding:2013hpa}
G.-J. Ding, S.~F. King, C.~Luhn, and A.~J. Stuart, ``{Spontaneous CP violation
  from vacuum alignment in $S_4$ models of leptons},''
  \href{http://dx.doi.org/10.1007/JHEP05(2013)084}{{\em JHEP} {\bfseries 05}
  (2013) 084},
\href{http://arxiv.org/abs/1303.6180}{{\ttfamily arXiv:1303.6180 [hep-ph]}}.

\bibitem{Ma:2006km}
E.~Ma, ``{Verifiable radiative seesaw mechanism of neutrino mass and dark
  matter},'' \href{http://dx.doi.org/10.1103/PhysRevD.73.077301}{{\em Phys.
  Rev.} {\bfseries D73} (2006) 077301},
\href{http://arxiv.org/abs/hep-ph/0601225}{{\ttfamily arXiv:hep-ph/0601225
  [hep-ph]}}.

\bibitem{Hirsch:2013ola}
M.~Hirsch, R.~A. Lineros, S.~Morisi, J.~Palacio, N.~Rojas, and J.~W.~F. Valle,
  ``{WIMP dark matter as radiative neutrino mass messenger},''
  \href{http://dx.doi.org/10.1007/JHEP10(2013)149}{{\em JHEP} {\bfseries 10}
  (2013) 149},
\href{http://arxiv.org/abs/1307.8134}{{\ttfamily arXiv:1307.8134 [hep-ph]}}.

\bibitem{Gu:2007gy}
P.-H. Gu, ``{Unified mass origin at TeV for dark matter and Dirac neutrinos},''
  \href{http://dx.doi.org/10.1016/j.physletb.2008.02.030}{{\em Phys. Lett.}
  {\bfseries B661} (2008) 290--294},
\href{http://arxiv.org/abs/0710.1044}{{\ttfamily arXiv:0710.1044 [hep-ph]}}.

\bibitem{Kanemura:2017haa}
S.~Kanemura, K.~Sakurai, and H.~Sugiyama, ``{Neutrino mass without lepton
  number violation, dark matter; and a strongly first-order phase
  transition},'' \href{http://dx.doi.org/10.1103/PhysRevD.96.095024}{{\em Phys.
  Rev.} {\bfseries D96} no.~9, (2017) 095024},
\href{http://arxiv.org/abs/1705.07040}{{\ttfamily arXiv:1705.07040 [hep-ph]}}.

\bibitem{Restrepo:2013aga}
D.~Restrepo, O.~Zapata, and C.~E. Yaguna, ``{Models with radiative neutrino
  masses and viable dark matter candidates},''
  \href{http://dx.doi.org/10.1007/JHEP11(2013)011}{{\em JHEP} {\bfseries 11}
  (2013) 011},
\href{http://arxiv.org/abs/1308.3655}{{\ttfamily arXiv:1308.3655 [hep-ph]}}.

\bibitem{Fu:2016ega}
{\bfseries PandaX-II} Collaboration, C.~Fu {\em et~al.}, ``{Spin-Dependent
  Weakly-Interacting-Massive-Particle-Nucleon Cross Section Limits from First
  Data of PandaX-II Experiment},''
  \href{http://dx.doi.org/10.1103/PhysRevLett.120.049902,
  10.1103/PhysRevLett.118.071301}{{\em Phys. Rev. Lett.} {\bfseries 118} no.~7,
  (2017) 071301}, \href{http://arxiv.org/abs/1611.06553}{{\ttfamily
  arXiv:1611.06553 [hep-ex]}}.
[Erratum: Phys. Rev. Lett.120,no.4,049902(2018)].

\bibitem{Aprile:2017iyp}
{\bfseries XENON} Collaboration, E.~Aprile {\em et~al.}, ``{First Dark Matter
  Search Results from the XENON1T Experiment},''
  \href{http://dx.doi.org/10.1103/PhysRevLett.119.181301}{{\em Phys. Rev.
  Lett.} {\bfseries 119} no.~18, (2017) 181301},
\href{http://arxiv.org/abs/1705.06655}{{\ttfamily arXiv:1705.06655
  [astro-ph.CO]}}.

\bibitem{LopezHonorez:2006gr}
L.~Lopez~Honorez, E.~Nezri, J.~F. Oliver, and M.~H.~G. Tytgat, ``{The Inert
  Doublet Model: An Archetype for Dark Matter},''
  \href{http://dx.doi.org/10.1088/1475-7516/2007/02/028}{{\em JCAP} {\bfseries
  0702} (2007) 028},
\href{http://arxiv.org/abs/hep-ph/0612275}{{\ttfamily arXiv:hep-ph/0612275
  [hep-ph]}}.

\bibitem{Babu:2002uu}
K.~S. Babu and C.~Macesanu, ``{Two loop neutrino mass generation and its
  experimental consequences},''
  \href{http://dx.doi.org/10.1103/PhysRevD.67.073010}{{\em Phys. Rev.}
  {\bfseries D67} (2003) 073010},
\href{http://arxiv.org/abs/hep-ph/0212058}{{\ttfamily arXiv:hep-ph/0212058
  [hep-ph]}}.

\bibitem{AristizabalSierra:2006gb}
D.~Aristizabal~Sierra and M.~Hirsch, ``{Experimental tests for the Babu-Zee
  two-loop model of Majorana neutrino masses},''
  \href{http://dx.doi.org/10.1088/1126-6708/2006/12/052}{{\em JHEP} {\bfseries
  12} (2006) 052},
\href{http://arxiv.org/abs/hep-ph/0609307}{{\ttfamily arXiv:hep-ph/0609307
  [hep-ph]}}.

\bibitem{Nebot:2007bc}
M.~Nebot, J.~F. Oliver, D.~Palao, and A.~Santamaria, ``{Prospects for the
  Zee-Babu Model at the CERN LHC and low energy experiments},''
  \href{http://dx.doi.org/10.1103/PhysRevD.77.093013}{{\em Phys. Rev.}
  {\bfseries D77} (2008) 093013},
\href{http://arxiv.org/abs/0711.0483}{{\ttfamily arXiv:0711.0483 [hep-ph]}}.

\bibitem{Alloul:2013raa}
A.~Alloul, M.~Frank, B.~Fuks, and M.~Rausch~de Traubenberg, ``{Doubly-charged
  particles at the Large Hadron Collider},''
  \href{http://dx.doi.org/10.1103/PhysRevD.88.075004}{{\em Phys. Rev.}
  {\bfseries D88} (2013) 075004},
\href{http://arxiv.org/abs/1307.1711}{{\ttfamily arXiv:1307.1711 [hep-ph]}}.

\bibitem{Babu:2016rcr}
K.~S. Babu and S.~Jana, ``{Probing Doubly Charged Higgs Bosons at the LHC
  through Photon Initiated Processes},''
  \href{http://dx.doi.org/10.1103/PhysRevD.95.055020}{{\em Phys. Rev.}
  {\bfseries D95} no.~5, (2017) 055020},
\href{http://arxiv.org/abs/1612.09224}{{\ttfamily arXiv:1612.09224 [hep-ph]}}.

\bibitem{Chatrchyan:2012ya}
{\bfseries CMS} Collaboration, S.~Chatrchyan {\em et~al.}, ``{A search for a
  doubly-charged Higgs boson in $pp$ collisions at $\sqrt{s}=7$ TeV},''
  \href{http://dx.doi.org/10.1140/epjc/s10052-012-2189-5}{{\em Eur. Phys. J.}
  {\bfseries C72} (2012) 2189},
\href{http://arxiv.org/abs/1207.2666}{{\ttfamily arXiv:1207.2666 [hep-ex]}}.

\bibitem{ATLAS:2012hi}
{\bfseries ATLAS} Collaboration, G.~Aad {\em et~al.}, ``{Search for
  doubly-charged Higgs bosons in like-sign dilepton final states at
  $\sqrt{s}=7$ TeV with the ATLAS detector},''
  \href{http://dx.doi.org/10.1140/epjc/s10052-012-2244-2}{{\em Eur. Phys. J.}
  {\bfseries C72} (2012) 2244},
\href{http://arxiv.org/abs/1210.5070}{{\ttfamily arXiv:1210.5070 [hep-ex]}}.

\end{thebibliography}\endgroup

\end{document}